\documentclass{emulateapj}
\usepackage{hyperref}
\usepackage{color}

\usepackage{grffile}
\newcommand{\msun}{M$_{\odot}$\,}

\newcommand\na{\ref@jnl{New A}}%

\defcitealias{do2012}{Paper I}

\shorttitle{Galactic Center IMF}
\shortauthors{Lu et al.}

\begin{document}

\title{Stellar Populations in the Central 0.5 pc of the Galaxy II:
The Initial Mass Function}

\author{
J. R. Lu\altaffilmark{1},
T. Do\altaffilmark{2},
A. M. Ghez\altaffilmark{3,4},
M. R. Morris\altaffilmark{3},
S. Yelda\altaffilmark{3},
K. Matthews\altaffilmark{5}
}

\altaffiltext{1}{Institute for Astronomy,
University of Hawaii, Honolulu, HI 96822;
jlu@ifa.hawaii.edu}

\altaffiltext{2}{Department of Physics and Astronomy,
University of California, Irvine, CA 92697;
tuan.do@uci.edu}

\altaffiltext{3}{Department of Physics and Astronomy,
University of California, Los Angeles, CA 90095-1547;
ghez@astro.ucla.edu, morris@astro.ucla.edu}

\altaffiltext{4}{Institute of Geophysics and Planetary Physics,
University of California, Los Angeles, CA 90095-1565}

\altaffiltext{5}{Division of Physics, Mathematics, and Astronomy,
California Institute of Technology, MC 301-17, Pasadena, CA 91125;
kym@caltech.edu}

\begin{abstract}
The supermassive black hole at the center of the Milky Way plays host
to a massive, young cluster that may have formed in one of the most
inhospitable environments in the Galaxy. We present new measurements
of the global properties of this cluster, including the
initial mass function (IMF), age, and cluster mass. These results are
based on Keck laser-guide-star adaptive optics observations
used to identify the young stars and measure their Kp-band luminosity function
as presented in \citet{do2012}.
A Bayesian inference methodology is developed
to simultaneously fit the global properties of the cluster utilizing
the observations and extensive simulations of synthetic star
clusters.
We find that the slope of the mass function for this cluster is
$\alpha = 1.7 \pm 0.2$, which is steeper than previously reported,
but still flatter than the traditional Salpeter slope of $2.35$.
The age of the cluster is between 2.5-5.8 Myr with 95\% confidence,
which is a younger age than typically
adopted but consistent within the uncertainties of past measurements.
The exact age of the cluster is difficult to determine since our results
show two distinct age solutions (3.9 Myr and 2.8 Myr) due to
model degeneracies in the relative number of Wolf-Rayet and OB stars.
The total cluster mass is between 14,000 - 37,000 \msun above 1 \msun and it
is necessary to include multiple star systems in order to fit the
observed luminosity function and the number of observed Wolf-Rayet stars.
The new IMF slope measurement is now consistent with X-ray observations
indicating a factor of 10 fewer X-ray emitting pre-main-sequence stars
than expected when compared with a Salpeter IMF.
The young cluster at the Galactic center is one of the few definitive
examples of an IMF that deviates significantly from the near-universal IMFs
found in the solar neighborhood.

\end{abstract}


\section{Introduction}
\label{sec:intro}
Young nuclear star clusters have now been found surrounding
supermassive black holes in a number of nearby galaxies
\citep[e.g.][]{lauer98,bender05,seth06}.
The best studied young nuclear cluster is at the
center of our own Milky Way Galaxy, located only 8 kpc away
and surrounding a supermassive black hole (SMBH) of mass M$\sim4 \times 10^6$ \msun
\citep{eckart97,ghez98pm,ghez00nat,ghez03spec,ghez05orbits,ghez08,
genzel00,schodel02,schodel03,eisenhauer06,gillessen2009}.
The young stars are located within 1 pc of the SMBH and are thought to have formed
as recently as 4-8 Myr ago \citep{paumard06}.
The origin of the young stars, in such close proximity to the supermassive
black hole, is puzzling, given that the strong tidal
forces in this region will shear apart typical molecular clouds
before they can collapse to form stars.
Thus, the young nuclear cluster (YNC) is a key laboratory for understanding
whether and how stars form under such extreme conditions.

The YNC at the Galactic center has several
observed properties that may help to determine its origin.
To date, more than 100 young stars have been spectroscopically identified
as OB supergiants, Wolf-Rayet stars, and, more recently, OB main-sequence
stars \citep{allen90,krabbe91,blum95heI,krabbe95,tamblyn96,
najarro97,ghez03spec,paumard06,bartko2010,do2012}.
The young population appears to fall into three dynamical categories;
(1) a well-defined clockwise rotating disk ranging
from 0.03 pc to at least 0.5 pc with moderate eccentricities,
(2) an off-the-disk population in a more isotropic
distribution over the same distances and also with moderate eccentricities, and
(3) $\sim$10\% cluster within 0.03 pc of the black hole with high
eccentricities of $\bar{e} = 0.8$
\citep{genzel00,levin03,genzel03cusp,paumard06,ghez08,lu09yng,gillessen2009b,yeldaThesis}.
These kinematic differences have led to some speculation that the young
stars may have been formed in multiple episodes, although the outer two groups (\#1 and \#2)
have consistent stellar populations \citep{paumard06}.
The inner group (\#3) was most likely dynamically injected, since the SMBH's tidal
forces are too strong to permit star formation at these close distances.
Initial theories suggested that this inner population was substantially older, as
long times were needed to scatter inward a large number of binary systems that
can interact with the SMBH and leave behind a star on a highly
eccentric orbit with a small semi-major axis \citep[e.g.][]{brown2007,perets2007}.
However, more recent inclusion of additional dynamical effects suggests
that the stars in the central region can be brought in more efficiently than
initially thought \citep{lockmann2009,madigan2011}.
Thus, it remains uncertain whether the inner group of young stars was originally
part of the outer groups and born in the same star formation event.
In the outer groups, there is some observational support for a possible warp in
the clockwise disk \citep{bartko09}, a second face-on disk made of counter-clockwise
rotating stars \citep{genzel03cusp,bartko09}, and for sub-clusters of
stars both on and off the disk
\citep{maillard04irs13,schodel05,lu05irs16sw}, although the statistical significance
of these results is still debated \citep{lu09yng,yeldaThesis}.
The total surface density profile of the young stars in the plane of the sky
is $\Sigma \propto R^{-1}$ \citep{do2012} and substantially steeper
in the disk plane $\Sigma_{\textrm{CW disk}} \propto R^{-2}$.
Early analysis of the bright stars (K$\leq$13) gave an age for the
cluster of 6 $\pm$ 2 Myr based on the presence of Wolf-Rayet (WR)
stars and the proportions of WR stars to O stars \citep{paumard06}.
In this same work, the observed number of WR and O stars and their
luminosity function suggested that the the total cluster mass is
$10^4$ \msun and the initial mass function (IMF, $dN/dm \propto
m^\alpha$) is top-heavy with a slope significantly flatter than
\citet[][$\alpha=2.35$]{salpeter1955}.
However, these results were limited by the lack of sensitivity to less massive,
main-sequence stars (K$>$13, M$<$20 \msun).
Deeper spectroscopic studies initially showed that the luminosity
function for the young stars appears to have a sharp turn-over at
K$=$13.5, suggesting that the present-day mass function is extremely
top-heavy with a slope of $\alpha=0.45 \pm 0.3$ \citep{bartko2010}.
A top-heavy IMF is also supported by the Chandra X-ray observations of the region,
given that lower mass young stars should still be coronally active and emitting
copious X-rays that aren't detected \citep{nayakshinSunyaev06}.

Several possible models for the origin of the young stars have been proposed and
the two presently supported by observations include
(1) formation {\em in situ} in a massive self-gravitating
molecular disk \citep{levin03} or (2) disruption of an in-falling massive cluster formed
much further away \citep{gerhard01}.
{\em In situ} formation models invoke the build-up of a gas disk surrounding
the SMBH that reached a critical mass ($\sim 10^4$ \msun) approximately
6 Myr ago such that local self-gravity within the disk became sufficient to overcome
the tidal shear, the disk collapsed along the
vertical direction and formed stars
\citep{levin03,kolykhalov80,shlosman89,morris96,sanders98,goodman03,nayakshinCuadra05}.
If such a disk was built up slowly, the gas and resulting young stars
would largely be on circular orbits, which does not appear to be the
case observationally \citep{lu09yng,bartko09}.
Modified {\em in situ} formation models include the rapid infall of
a single massive molecular cloud or the collision of two in-falling clouds
to produce non-circular orbits and the on-disk and off-disk populations
\citep{alexander07imf,cuadra08,sanders98,vollmerDuschl01,nayakshin07sims,hobbs2009}.
The surface density of stars formed {\em in situ} in a slowly-built gas disk
may be steep ($\Sigma \propto R^{-2}$); however, the surface
density resulting from a cloud-cloud collision is less well constrained.
For in-falling cluster scenarios, the young stars are formed a few
parsecs away in a massive young star cluster that spirals in via dynamical
friction and disrupts in the central parsec \citep{gerhard01}.
In order for such a cluster to migrate into the central parsec within
only a few million years, the cluster must be very massive and centrally
concentrated and perhaps even host an intermediate mass black hole (IMBH)
at its center \citep{kim03,pzwart03irs16,mcmillan03,gurkan05,hansen03,kim04}.
Such a scenario would produce the on-disk population and a flatter
surface density profile ($\Sigma \propto R^{-0.75}$), with the
off-disk population resulting from subsequent dynamical perturbations
\citep{haas2011,baruteau2011}.
{\em In situ} formation models are favored based on the density
profiles and available time scales;
however, no single model completely explains all of the observed properties.
One important prediction of current {\em in situ} formation models is that the
initial mass function should be very top-heavy due to the extreme temperatures,
pressures, densities, and ambient radiation fields present in the central
parsec \citep{nayakshin06thick,nayakshin2006imf,alexander07imf,alexander08,cuadra08}.
However, a number of model parameters are still very uncertain
(e.g. gas infall rate, temperature, pressure, and cooling time)
and a wide variety of initial mass functions are still possible.

The initial mass function, age, and mass of the young nuclear star
cluster at the Galactic center is of particular interest both for
understanding the origin of the young stars and exploring star formation under extreme conditions.
Past observations of a turn-over in the near-infrared luminosity function and the low
total X-ray emission apparently support a top-heavy IMF.
However, new observations presented in our companion paper,
\citet[][Paper I]{do2012}, show a luminosity function that does not turn over,
but continues to rise; warranting a new analysis of the mass function.

In this work, we compare the near-infrared photometry of the
spectroscopically identified young stars in Paper I to models of star
clusters in order to determine the age, mass function, and total mass
of the young nuclear star cluster.
Details on the observational sample and measurements are presented in \S\ref{sec:obs_and_analysis}.
Synthetic clusters are generated using
stellar evolution and atmosphere models to produce individual stars as they
would appear at the Galactic center in \S\ref{sec:synthetic_clusters}.
In \S\ref{sec:bayesian_methodology} we present a Bayesian inference method
for fitting the observed data with synthetic clusters to determine the
cluster's properties and uncertainties.
Results in \S\ref{sec:results} show that the present-day mass function
is slightly flatter than a Salpeter IMF
and rule out a mass function as top-heavy as previously claimed.
We also find a younger age and our modeling requires the inclusion of binaries
and multiple star systems to adequately fit the luminosity function.

\section{Observed Data \& Sample}
\label{sec:obs_and_analysis}

\begin{figure}
\begin{center}
\includegraphics[scale=0.5]{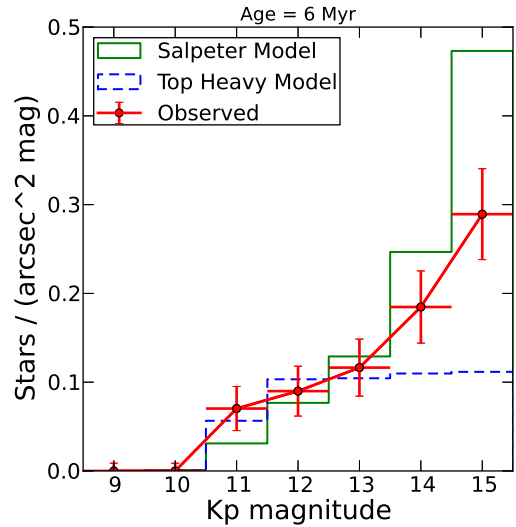}
\end{center}
\caption{The observed Kp luminosity function (KLF) for early-type stars in the Galactic center nuclear cluster (Paper I) compared with previously proposed model KLFs.
The observed Kp magnitudes ({\em red}) are corrected for differential extinction to a common extinction value of A$_{Ks}=2.7$.
A model KLF is shown for published best-fitting cluster parameters ({\em blue}) with an IMF slope of $\alpha=0.45$, age of 6 Myr, and a distance of 8 kpc \citep{bartko2010}.
For comparison, a model KLF is also shown for a similar cluster with a Salpeter IMF slope of $\alpha=2.35$ ({\em green}).
}
\label{fig:klf}
\end{figure}

Observations and data used in this paper are described in
\citetalias{do2012}.
Our data sample includes all stars with a
non-zero probability of being a young, early-type star ($p_{yng} > 0$).
For each star, the measurements consist of a Kp magnitude
corrected for differential extinction (K$^\prime_{\Delta A}$),
a Kp uncertainty ($\sigma_{K^\prime}$), the probability of youth
($p_{yng}$), and an indicator for whether the star is a Wolf-Rayet star.
Only stars with K$^\prime_{\Delta A} \leq 15.5$ are included.
The modeling presented in this work utilizes these individual
measurements rather than a binned Kp luminosity function (KLF).
However for illustrative purposes, Figure \ref{fig:klf} shows our binned
KLF compared with a model KLF for a cluster with an age of 6 Myr,
extinction of A$_{Ks}=2.7$, distance of 8 kpc, and an IMF slope of
$\alpha=0.45$, consistent with previously published best-fit cluster
parameters \citep{bartko2010}.
The discrepancy between our observed KLF and this model motivates the
work presented below as we attempt to derive new best-fit
cluster parameters and uncertainties.

In our sample, we have chosen to include young stars at all radii, including
the central 0\farcs8 region immediately around the supermassive black hole.
There are some suggestions that the stars in this central region
were not formed in the most recent starburst, but rather in an earlier
event.
The brightest early-type star in this region, S0-2, is a B0-B2.5 V star
with an age of less than 15 Myr based on its measured temperature, gravity,
and luminosity \citep{martins08}.
The other early-type stars in this region have spectral features consistent with
main-sequence stars \citep{eisenhauer06} and the oldest age for a main-sequence
star at Kp$=$15.5 (A$_{Ks}=2.7$, d$=$8 kpc) is less than 20 Myr.
Given these age constraints and the large uncertainties in the theoretical
scattering efficiencies, it remains plausible that the young stars in the central region
of our sample are the same age as the stars at larger radii.
Furthermore, as described in \citetalias{do2012}, including or discarding the stars within
a radius of 0\farcs8 does not significantly change the shape of the KLF.

\section{Modeling the Observed Cluster}
\label{sec:models}

While the early-type stellar population in the central parsec of the Galaxy can be divided
into several different kinematic sub-groups \citep{paumard06,ghez08,lu09yng,bartko09,gillessen2009b}, it is not yet clear
whether these sub-groups have different origins, or whether all the young stars formed from
the same starburst and were differentiated via subsequent dynamical processes.
Therefore, we choose to analyze the entire population of young stars as a single starburst
cluster and to estimate the cluster's mass function, age, and total mass.

The traditional approach to deriving a mass function would be to construct a binned
KLF from the observed Kp magnitudes, assume a distance and extinction, and use a mass-luminosity
relation from models of stellar evolution and atmospheres to convert
from observed magnitudes to initial stellar masses.
This approach has several shortcomings, including arbitrarily choosing
bin sizes, neglecting uncertainties on stellar brightness, not accounting for multiple systems,
and correlations in the cluster age, mass, and IMF
\citep{maizApellaniz2005,maizApellaniz2008,maizApellaniz2009}.
More recent and rigorous statistical methods, particularly those using Bayesian inference,
provide flexibility and produce robust estimates of the most probable cluster parameters
and their uncertainties \citep[e.g.][]{allen2005}.
We have developed a Bayesian inference methodology for deriving the physical properties of
an observed young star cluster using simulations of synthetic clusters.
The process of simulating a synthetic young cluster is described in \S \ref{sec:synthetic_clusters}
and the full Bayesian methodology, implementation,
and testing is described in \S \ref{sec:bayesian_methodology}, \S \ref{sec:multinest},
and \S \ref{sec:test_sim_clusters}, respectively.
Readers wishing to skip the detailed description of the modeling may proceed to
\S \ref{sec:results} where the resulting best-fit cluster properties and uncertainties
are presented.

\subsection{Synthetic Clusters}
\label{sec:synthetic_clusters}
Model young star clusters are produced assuming an instantaneous starburst at time $t$,
at distance $d$, with an observed cluster mass $M_{cl,obs}$.
The initial mass function of the cluster is described as a single power-law given by the
probability density function
\begin{eqnarray}
\frac{dN}{dm} \propto p(m) & = &  \frac{(1-\alpha) m^{-\alpha}}{m_{max}^{1-\alpha} - m_{min}^{1-\alpha}}
\quad [\alpha \neq 1] \\
& = & \frac{m^{-\alpha}}{\ln m_{max} - \ln m_{min}} \quad [\alpha = 1]
\end{eqnarray}
where $\alpha$ is allowed to vary and has the value $\alpha=2.35$ in the case of a typical
IMF for the high-mass end of a solar-neighborhood population \citep{salpeter1955}.
In addition to these free parameters, a number of fixed parameters are
also needed to fully model the cluster.
First, the extinction is fixed to A$_{Ks}=2.7$ as we will compare to
observations that have been corrected for differential extinction to
this value.
The metallicity is fixed to Z=0.02, which is roughly solar.
Observations suggest the young GC population is consistent with solar
iron abundance; but may be as high as 2$\times$ solar and seems enhanced in
$\alpha$-elements \citep{carr2000,ramirez2000,najarro2004,cunha2007,martins08}.
The choice to keep metallicity fixed to solar is also motivated by the lack of complete
stellar evolution and atmosphere models at other metallicities and abundance ratios.
Section \ref{sec:discussion} discusses possible uncertainties associated with
restricting our analysis to solar metallicity.
The range of masses we consider is set as $m_{min}$ = 1 \msun and $m_{max}$ = 150 \msun
and is kept fixed throughout the analysis.
This is justified as our observations do not include stars below 10 \msun
due to sensitivity limits and stars above 150 \msun have
already exploded for clusters older than 3 Myr.
Thus the currently observed stars have no power to constrain the low
or high mass cutoffs and a more expansive mass range would not impact our results.
For each star drawn from the IMF, we must also consider whether it has
companions \citep{sagar1991,kroupa1995,goodwin2005,thies2007,maizApellaniz2009,weidner2009}.
High mass stars in the Galactic disk are known to have a high multiplicity fraction, perhaps
even 100\% \citep{kobulnickyFryer2007} and the multiplicity fraction and the
companion star fraction, which is the mean number of companions per primary, are dependent on
the mass of the primary star in the multiple system
\citep[e.g.][]{lafreniere2008}.
The Galactic center cluster is modeled allowing multiple systems with a mass-dependent
multiplicity fraction (MF) and companion star fraction (CSF).
The functional form of how the MF and CSF vary with primary mass
is still very uncertain even for solar-neighborhood young clusters.
We empirically derive this relation by compiling MF and CSF measurements in young
clusters from the literature and assuming that the MF and CSF follow a
power law distribution (see Appendix \ref{sec:multiplicity} for the derivation).
Each star drawn from the IMF with mass $m$ is determined to be
in a multiple system and assigned a number of companions based on
\begin{eqnarray}
MF(m) &=& 0.44 \; \left(\frac{m}{1 \;M_{\odot}}\right)^{0.52} \;\;\;\; \textrm{always} \leq 1 \\
CSF(m) &=& 0.48 \; \left(\frac{m}{1 \;M_{\odot}}\right)^{0.49} \;\;\;\; \textrm{always} \leq 3.
\end{eqnarray}
The mass of the companion stars is drawn from a power law probability
distribution on the mass ratio, $q$:
\begin{eqnarray}
p(q) = \left( \frac{1 + \beta}{1 - q_{lo}^{1+\beta}} \right) q^\beta
\end{eqnarray}
where $\beta = -0.4$ and the lowest allowed mass ratio is $q_{lo} = 0.01$.
Our choice of $\beta$ is consistent with observations of local massive
stars in young OB associations, although values ranging from
$\beta=0.4 - -0.6$ would also be within 68\% confidence intervals
allowed by observations\footnote{The resulting companion star masses
are substantially higher than randomly drawing companions from the IMF.}
\citep{kobulnickyFryer2007,kiminki2012}.
Luminosity functions are hardly impacted by this choice of
$q_{lo}$ since lower mass ratios do not significantly increase the
total system luminosity \citep{kouwenhoven2009}.
Allowing for multiple systems will have the largest effect at the bright
end of the luminosity function where magnitude bins that were un-populated
by individual stars are filled in with multiple star systems composed
of massive, bright primaries and companions \citep{maizApellaniz2009,weidner2009}.
In \S\ref{sec:test_sim_clusters}, simulated clusters are used to show
that incorporating multiple systems is essential and that failing to
include them produces poor fits and incorrect ages and IMF slopes.
For a cluster with a desired cluster mass of $M_{cl,obs}$, stars are simulated until
the sum of all stellar masses, including companions, exceeds $M_{cl,obs}$ and then the
last simulated star is thrown out if that would bring the synthesized
cluster mass closer to the desired $M_{cl,obs}$ \citep{haas2010thesis}.
A complete list of free and fixed parameters we use to produce synthetic star clusters
is presented in Table \ref{tab:model_parameters}.

\begin{figure}
\begin{center}
\includegraphics[scale=0.5]{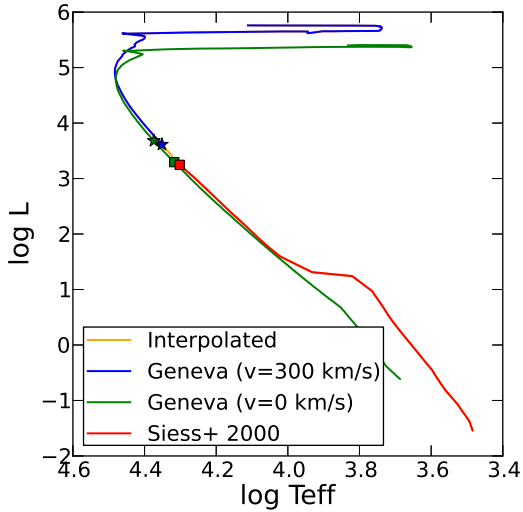}
\end{center}
\caption{
6 Myr luminosity-effective temperature isochrones from various stellar evolution models
used to construct and test a combined grid of stellar evolution models that spans ages
from 1-25 Myr and the full range of stellar masses.
Geneva models without rotation spanning 0.8$-$120 \msun are shown in {\em green} as a
baseline \citep{schaller1992}.
The observed post-main-sequence evolution of massive stars at this age is best described
with models that include rotation.
The isochrone from Geneva models with rotation, shown in {\em blue}, differs significantly
in luminosity.
Models with rotation only extend down to 9 \msun ({\em blue star} and {\em green star}).
Models provided by \citet{siess00} are used for main sequence and pre-main sequence
models below 7 \msun and are shown in {\em red} (7 \msun point also shown
as a {\em red square} and {\em green square}).
The {\em blue} and {\em red} isochrones are merged and interpolated over the gap to
construct final isochrones, shown in {\em orange}, used in our cluster modeling.
}
\label{fig:merge_evo_models}
\end{figure}

Conversion from stellar masses to synthetic photometry requires the
use of stellar evolution and atmosphere models.
At the suspected age of the young cluster (4$-$8 Myr), stars with masses
in excess of 30-60 \msun are evolving off the main-sequence and
stellar evolution models with
rotation are found to best describe populations of blue supergiants
and Wolf-Rayet stars \citep{meynet2003}.
We use the grid of Geneva models with an initial rotation speed of 300
km/s provided by \citet{meynet2003}, which only extends down
to 9 \msun\footnote{\citet{ekstrom2012} contains an updated grid of Geneva models with rotation
extending down to lower masses and with rotation speeds other than 300
km/s; however, they were not available at the time of our analysis.}.
In these models, Wolf-Rayet stars are identified using
the criteria suggested in \citet{meynet2003}: $T_{eff}>10^4$ K and the mass fraction of
hydrogen at the surface is $X_s<0.4$.
To extend to lower masses along the main sequence and also capture pre-main sequence evolution at
younger ages, the Geneva grid is merged with the grid of models provided by \citet{siess00} that
extends up to 7 \msun.
For both grids, a suite of isochrones is generated at logarithmic ages of
$\log t = 6.0-7.4$ with steps of $\Delta \log t = 0.01$ and each isochrone
samples mass more finely than $\Delta m = 0.005$ \msun.
The two isochrones are then merged and interpolated over the gap between the two models (7-9 \msun).
An example of the merging process for $\log t=6.8$ is shown in
Figure \ref{fig:merge_evo_models}.
The resulting isochrones provide a mapping from the simulated stellar masses to their
luminosities ($L$), effective temperatures ($T_{eff}$) and surface gravities ($g$).

The physical parameters for each star are converted into observable brightness at Kp-band using
model stellar atmospheres.
Again, there is no single set of atmosphere models that spans the full range of effective
temperatures and surface gravities.
For hot stars with $T_{eff} > 7000$ K, atmosphere models by \citet{castelli2003} are used
for all non-Wolf-Rayet stars.
For temperatures of 4000 K $< T_{eff} < $ 7000 K, we use the NextGen models of \citet{hauschildt1999}
and for $T_{eff} < $ 4000 K we use the updated NextGen models with improved AMES opacities
\citep{allard2000}.
The grids do not intersect exactly at the transition temperatures; however,
synthetic broadband photometry differs by $\lesssim$2\% in the K-band, which is less
than our photometric precision, so no interpolation between the different atmosphere
models is done.
All downloaded atmosphere models were re-formatted to work with the {\em pysynphot}
python package in order to perform interpolations within each grid and generate
synthetic spectra at specific temperatures, metallicities, and gravities.
The synthetic spectra are flux calibrated to what would be observed at
Earth without extinction by multiplying by $(R/d)^2$,
where $R$ is the radius of the star and $d$ is the distance.
Then the synthetic spectra are reddened using the Galactic Center extinction law
from \citet{nishiyama09} and convolved with transmission profiles for a typical Mauna Kea
atmosphere and for the Kp filter in NIRC2.
The final synthetic spectra are integrated using {\em pysynphot} to produce
broad-band Kp photometry as would be seen from the telescope.
An example mass-Kp magnitude relation is shown in
Figure \ref{fig:klf_vs_multiples} (left panel) and would
predict that S0-2, with a Kp$=14.39$ after correcting to A$_{Ks}=2.7$, would have a
mass of $\sim$18 \msun, which is consistent with previous estimates based on the
effective temperature and surface gravity derived from detail spectral analysis
\citep{martins08}.

Wolf-Rayet (WR) stars require special consideration as their atmospheres are windy and clumpy.
Models of stellar evolution for WR stars can predict when stars enter and exit the WR phase,
surface abundances, and the luminosities and temperatures at the base of the atmosphere.
However, this is insufficient to uniquely define what will be observed due to differences in
wind velocities, mass loss rates, and clumping factors \citep[e.g.][]{hamann2008}.
Put another way, there is no unique mass-magnitude relationship for Wolf-Rayet stars.
Model atmospheres do exist for Wolf-Rayet stars; but they require observations of the
wind velocities and mass-loss rates and still do not uniquely predict the interior properties
of the star, such as temperature and gravity at the base of the atmosphere \citep{martins07}.
Therefore, the observed magnitudes of the Wolf-Rayet stars cannot be used in a straightforward
manner and we exclude them from the analysis of the luminosity function.
The number of WR stars, N$_{WR}$ relative to the number of OB stars is still a powerful
constraint on the age and mass of the cluster and we incorporate N$_{WR}$ into our Bayesian analysis.

\begin{deluxetable*}{lllp{2in}}
\tablewidth{0pt}
\tablecaption{Model Cluster Parameters}
\tablehead{
\colhead{Parameter} &
\colhead{Vary?} &
\colhead{Description} &
\colhead{Prior}
}
\startdata
$\alpha$ & free & IMF slope between $m_{min}$ and $m_{max}$ & Uniform from [0.10, 3.35] \\
$\log t$ & free & log$_{10}$( age of the young cluster in years ) & Gaussian with $\bar{\log t} = 6.78$, $\sigma = 0.18$ from [6.20, 6.70] \\
$M_{cl,obs}$ & free & initial mass of the young cluster & Uniform from [10$^3$, 10$^5$] \msun \\
d & free & distance to the young cluster & Gaussian with $\bar{d}=8.096$ kpc, $\sigma=0.483$ kpc from [6.793, 9.510] \\
$m_{min}$ & fixed & IMF minimum stellar mass & 1 \msun \\
$m_{max}$ & fixed & IMF maximum stellar mass & 150 \msun \\
Z & fixed & metallicity of the young cluster & 0.20 (roughly solar) \\
A$_{Ks}$ & fixed & extinction to the young cluster & 2.7 \\
MF\_amp & fixed & MF function's amplitude & 0.44 \\
MF\_index & fixed & MF function's power-law index & 0.51 \\
CSF\_amp & fixed & CSF function's amplitude & 0.50 \\
CSF\_index & fixed & CSF function's power-law index & 0.45 \\
CSF\_max & fixed & maximum value for CSF & 3 \\
$\beta$ & fixed & power-law index for $q$ distribution & -0.4 \\
\enddata
\label{tab:model_parameters}

\end{deluxetable*}

\subsection{Bayesian Methodology}
\label{sec:bayesian_methodology}

A Bayesian inference approach is used to determine the physical properties of the
YNC at the Galactic center. This methodology is generally applicable
to the analysis of other star clusters.
The advantages of Bayesian inference for astronomical applications are highlighted in
\citet{press1997} and have been widely adopted in cosmology \citep{hobson2010}.
Bayesian techniques are beginning to be used more widely in the study of all types of
star clusters
\citep{selman1999,allen2005,converseStahler2008,deGennaro2009,vanDyk2009,espinoza09,rizzuto2011}.
The analysis presented here differs from previous work in several ways.
First, single-band photometric information is combined with some spectroscopic
information that identify which stars are Wolf-Rayet stars.
In our case, only single band photometry is necessary since extinction maps are
available from past multi-band photometry \citep{schodel2010b}.
Second, the initial mass function is assumed to have the functional form of a
truncated power-law, although other functional forms can easily be inserted.
This differs from some Bayesian treatments that attempt to derive the masses of the
individual stars allowing for infinite variation in the shape of the mass function
\citep{converseStahler2008,vanDyk2009,deGennaro2009}.
Third, we allow for mass-dependent multiplicity fractions and high-order
multiple systems, which are important at the high mass end of the mass function.
Fourth, we include all candidate young stars and weight them by their probability
of youth in order to use all available information and accurately account
for incompleteness\footnote{A significant number of the candidate young stars
were spectral-typed manually and were assigned 100\% probability of youth.
See \citet{do2012} for a complete discussion.}.
Finally, instead of using Markov Chain Monte Carlo techniques to perform the Bayesian
inference, we find that better solutions result from nested sampling
techniques \citep{feroz2008,feroz2009}.
This Bayesian approach provides detailed probability distributions for the mass
function slope and other cluster parameters given limited observations.
It also avoids traditional biases from binning and neglecting photometric errors and
accounts for stochastic sampling of stellar masses.
Our full methodology is described below in more detail.

We start at the beginning, with Bayes equation
\begin{equation}
p(\Theta | D) = \frac{p(D | \Theta) p(\Theta)}{p(D)}
\end{equation}
where $D$ is the observed data, $\Theta$ is the model, $p(D | \Theta)$ is the
likelihood of observing the data given some model, $p(\Theta)$ captures the
prior knowledge on the model parameters, and $p(D)$ is a normalizing factor known
as the evidence.
The likelihood function and the prior distributions then give
posterior probability distributions for the model parameters, $p(\Theta | D)$.
The model, $\Theta$, is defined by the set of free and fixed parameters given in
Table \ref{tab:model_parameters}.

The data, $D$, is constructed from the sample of candidate young stars ($p_{yng} > 0$)
brighter than Kp$=$15.5 as described in \S\ref{sec:obs_and_analysis}.
The WR stars are separated out and only the observed number of WR stars, $N_{WR}$, is
included in our analysis, since the initial masses of WR stars cannot be determined
based on their luminosity alone.
For all the other non-WR stars, measurements of the Kp brightness and error are used
along with the probability that each star is young, $p_{yng}$.
Although we observe $N_{obs}$ non-WR stars, not all of them are young.
We estimate the number of young OB stars, $N_{OB}$, as the
sum of the observed non-WR stars, weighted by their probability of youth:
$N_{OB} = \sum_{i=1}^{N_{obs}} p_{yng,i}$.
The resulting measurements that make up the data are then
\begin{equation}
D = [N_{WR}, \mathbf{k_{obs}}, N_{OB}]
\end{equation}
where $\mathbf{k_{obs}}$ is the set of $\{k_i, \, \sigma_{k_i}, \, p_{yng,i}\}$
measurements for the $N_{obs}$ non-Wolf Rayet stars observed.

The likelihood function is composed of three independent probabilities:
(1) the likelihood of observing the set of Kp
magnitudes, $\{k_i\}$, and uncertainties, $\{\sigma_{k_i}\}$,
which effectively captures the shape of the KLF;
(2) the likelihood of observing $N_{OB}$ OB stars, which
captures the normalization of the KLF; and
(3) the likelihood of observing $N_{WR}$ Wolf-Rayet stars:
\begin{equation}
p(D | \Theta) =  p(\mathbf{k_{obs}} | \Theta)
\cdot p(N_{OB} | \Theta).
\cdot p(N_{WR} | \Theta).
\label{eqn:likelihood}
\end{equation}
To derive the first term, we use an unbinned approach such that the
likelihood is the multiplication of the individual stars' probabilities,
weighted by the probability of youth,
\begin{equation}
p(\mathbf{k_{obs}}|\Theta) = \prod_{i=1}^{N_{obs}} p(k_i|\Theta)^{p_{yng,i}}.
\label{eqn:likelihood_obs}
\end{equation}
The probability for a star to have some Kp, $p(k_i|\Theta)$, is
given by the probability distribution derived from
synthetically ``observing'' a simulated young cluster.
First, the cluster is simulated with model
parameters describing the age ($t$), IMF slope ($\alpha$),
cluster mass ($M_{cl,obs}$), and the cluster distance ($d$),
\begin{equation}
p(k|\Theta)_{intrinsic} = \textrm{Simulated Yng Cluster}(t, \alpha, M_{cl,obs}, d).
\end{equation}
These free parameters along with their priors are given in
Table \ref{tab:model_parameters}.
The resulting Kp photometry from the simulated stars in the model cluster are
binned finely (0.1 mag bins) to produce a probability distribution for
the intrinsic Kp luminosities.
For small numbers of stars, stochastic sampling effects can lead to an inaccurate
estimate of the probability distribution.
Therefore, in order to maximize accuracy and reduce the total number of clusters
that need to be simulated, all model young clusters are simulated with a total
mass of $10^6$ \msun.
The shape of the Kp luminosity function does not change with cluster mass and
the number of Wolf-Rayet and non-Wolf-Rayet stars simply scales linearly with total
cluster mass.
The intrinsic probability distribution for Kp luminosities is multiplied by
the imaging completeness curve, $C(k)$,
truncated at Kp$=$15.5 to match the data, and normalized to
give the final probability distribution,
\begin{equation}
p(k|\Theta)_{observed} = \frac{p(k|\Theta)_{intrinsic} \cdot C(k)}{
\int^{15.5}_0 p(k|\Theta)_{intrinsic} \cdot C(k) \; dk}.
\end{equation}
Finally, the measured photometric error is incorporated by modeling the
observed magnitude as
$k_i = k' + \epsilon$ where $\epsilon$ is drawn from a normal distribution
with zero mean and a standard deviation of
$\sigma_{k_i}$. The resulting probability of observing $k_i$ for a given star
is then given by
\begin{equation}
p(k_i | \Theta) = \int^\infty p(k' | \Theta)_{observed} \;
\frac{1}{\sqrt{2 \pi \sigma_{k_i}^2}} \; e^{\frac{-(k' - k_i)^2}{2
\sigma_{k_i}^2}} dk',
\end{equation}
which is the convolution of the model KLF and the error probability distribution.
The resulting probabilities for each star feed into Equation \ref{eqn:likelihood_obs}
to calculate the first term of the total likelihood in Equation \ref{eqn:likelihood}.
In the second and third terms of the likelihood, the number of OB and WR
stars scales linearly with the mass of the young cluster, which was verified
with simulations.
Thus the same simulated young clusters used to produce $p(k|\Theta)_{intrinsic}$
also give the expected number of WR and OB stars after scaling linearly with
cluster mass,
\begin{eqnarray}
\aleph_{WR}(M_{cl,obs}) = \aleph_{WR}(10^6 M_\odot) \left(\frac{ M_{cl,obs}}{10^6 M_\odot}\right) \\
\aleph_{OB}(M_{cl,obs}) = \aleph_{OB}(10^6 M_\odot) \left(\frac{ M_{cl,obs}}{10^6 M_\odot}\right)
\end{eqnarray}
where $\aleph_{WR}$ and $\aleph_{OB}$ are the expectation values (means) for the number of
WR and non-WR stars, respectively.
The likelihood of observing $N_{WR}$ and $N_{OB}$ are then taken as a Poisson distributions
\begin{eqnarray}
p(N_{WR} | \Theta) = \frac{\aleph_{WR}(\Theta)^{N_{WR}} \; e^{\aleph_{WR}(\Theta)}}{N_{WR}!} \\
p(N_{OB} | \Theta) = \frac{\aleph_{OB}(\Theta)^{N_{OB}} \; e^{\aleph_{OB}(\Theta)}}{N_{OB}!}.
\end{eqnarray}

\subsection{Sampling Posterior Probability Distributions with Multinest}
\label{sec:multinest}

The resulting posterior probability distribution for the model parameters cannot be
calculated analytically.
Traditionally, Monte Carlo techniques are used to produce a representative sample of
points from the posterior probability distribution.
The most commonly used method is the Markov Chain Monte Carlo (MCMC).
However, MCMC methods may have
difficulty converging or fully mapping probability space when the probability
distributions that are being sampled are multi-modal or highly degenerate and curved.
An alternative method is nested sampling \citep{skilling2004} such as in
the publicly available multi-modal nested sampling code,
{\em \href{http://ccpforge.cse.rl.ac.uk/gf/project/multinest/}{MultiNest}}
\citep{feroz2008,feroz2009}.
This algorithm has been successfully demonstrated in cosmology, galaxy evolution,
and gravitational wave problems \citep[e.g.][]{bridges2009,kilbinger2010,martinez2011,veitch2010}
and is less computationally expensive and more accurate, in some cases, than MCMC methods.
We performed tests on simulated clusters using traditional MCMC techniques using both
Metropolis-Hastings and Hit-and-Run step methods (\href{http://code.google.com/p/pymc/}{PyMC package}) and compared the
overall computation time and accuracy with the MultiNest code.
MultiNest took 5-10 times less computation time and produced more accurate results in all our
simulated clusters (see \S\ref{sec:test_sim_clusters} for more details).
Therefore, we utilized MultiNest in our analysis of the Young Nuclear Star cluster's
IMF.

MultiNest uses a fixed number of points per iteration to sample parameter space and calculate
the Bayesian evidence at each point.
In successive iterations, the same number of points are concentrated
into smaller and smaller volumes centered around the most probable regions.
This process continues until the evidence no longer changes by more than a specified
tolerance value.
We use 1000 points and an evidence tolerance of 0.3 in order to sample parameter space well
and run efficiently.
We also enabled multi-modal searches as several simulated clusters could clearly be fit
by several distinct sets of model parameters.

\subsection{Testing on Simulated Clusters}
\label{sec:test_sim_clusters}

To test the speed and accuracy of our bayesian methodology and codes, clusters were
simulated and synthetically  ``observed'' that had different ages, IMFs, and
multiplicity properties.
The synthetic ``observations'' were then fit with the bayesian inference techniques
described above and the resulting posterior probability distribution functions were
examined to see if they produced results consistent with the input cluster parameters.
All clusters were simulated at a distance of 8 kpc, an extinction of $A_{Ks}=2.7$,
and with a cluster mass aimed at producing a similar number of OB stars to our observed
data set.
Photometric errors for simulated cluster stars are randomly drawn from
the distribution of empirical errors from the GC observations.
Tests were performed to
(1) understand degeneracies in fitting a cluster's age,
(2) determine the impact of multiplicity, and
(3) explore whether we recover both top-heavy and normal IMF slopes.

\subsubsection{Age}

\begin{figure}
\begin{center}
\includegraphics[scale=0.5]{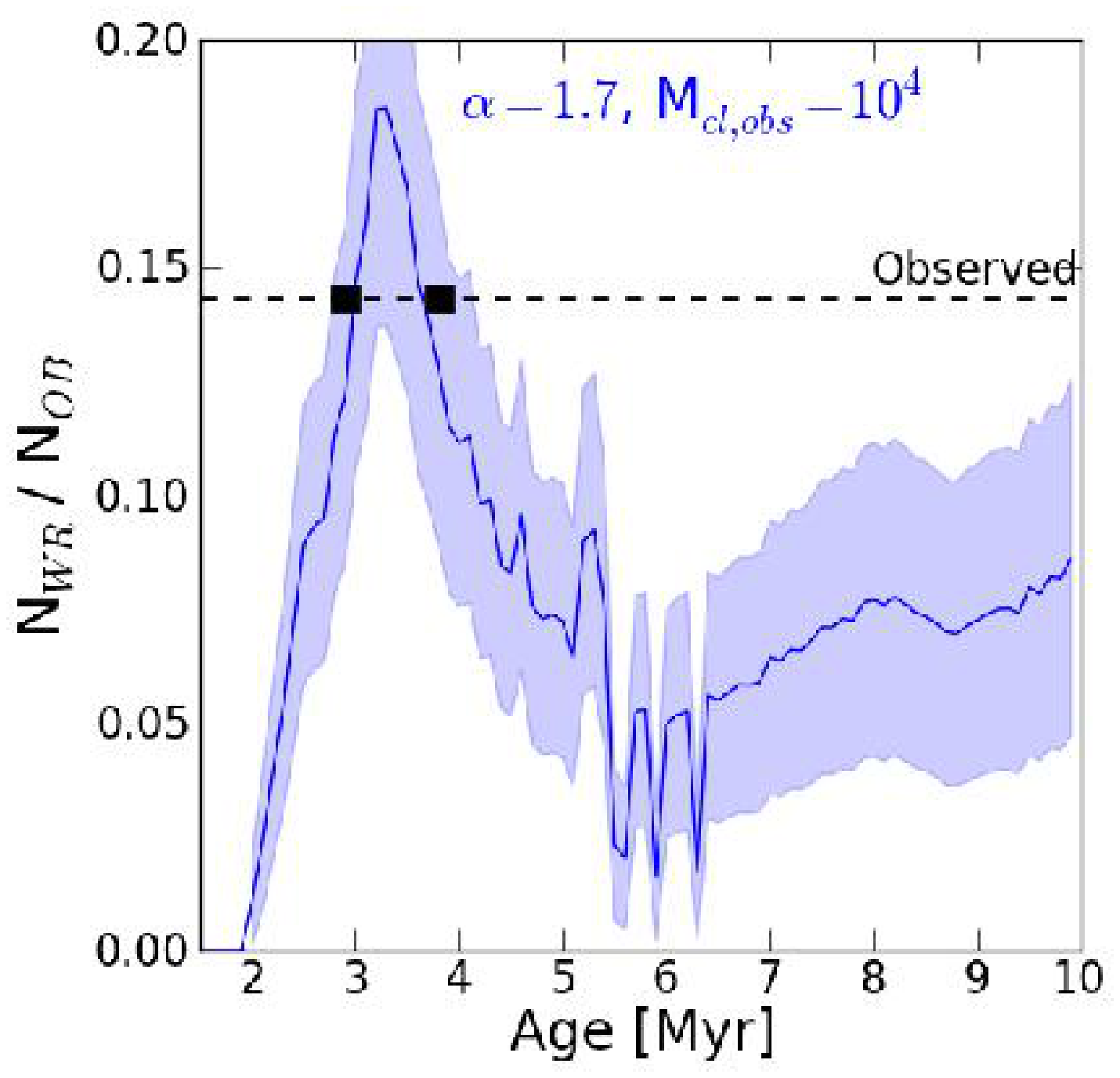}
\includegraphics[scale=0.5]{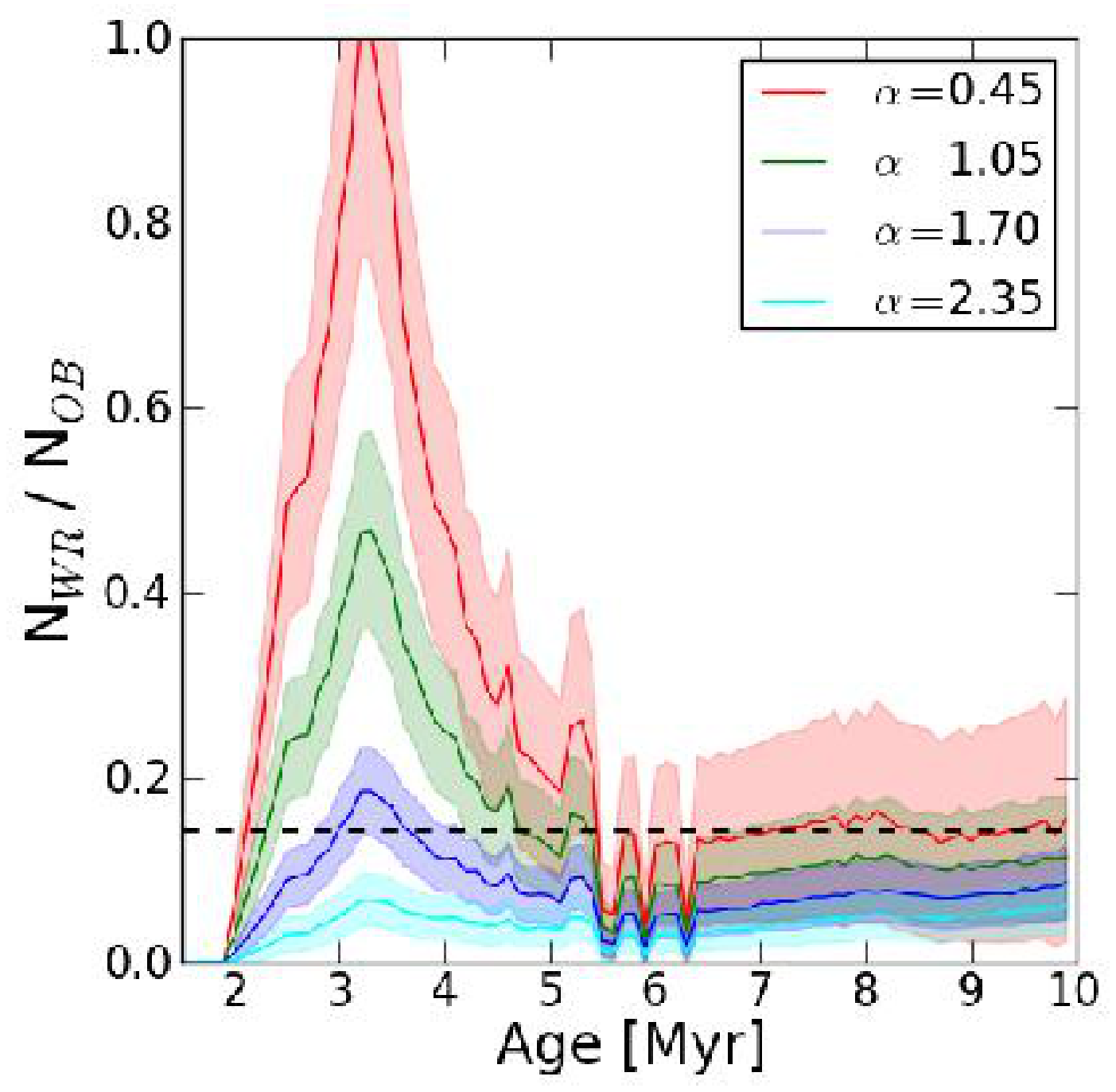}
\end{center}
\caption{
Ratio of the number of WR stars to the number of OB stars
(Kp$\leq$15.5) as a function of cluster age.
This quantity, along with the magnitude of the brightest observed OB stars,
provides the strongest constraint on the cluster age; however, it
shows degeneracies that often result in multiple solutions for a given cluster.
{\em Left:} An example cluster with an IMF slope of $\alpha=1.7$
will produce a specific N$_{WR}$/N$_{OB}$ ratio two to three times in
the first 10 Myr ({\em blue solid}).
{\em Right:} Changing the IMF slope will shift the ages at which a specific
N$_{WR}$/N$_{OB}$ ratio occurs.
The observed N$_{WR}$/N$_{OB}$ ratio for the YNC intersects the $\alpha=1.7$
cluster at ages of $\sim$3 and $\sim$4 Myr.
Uncertainties due to stochastic sampling ({\em shaded}) also suggest that 8-10 Myr
may be a plausible solution; but, the presence of bright O giants and
supergiants makes these larger ages unlikely.
}
\label{fig:age_vs_WR_OB}
\end{figure}

\begin{figure}
\begin{center}
\includegraphics[scale=0.4]{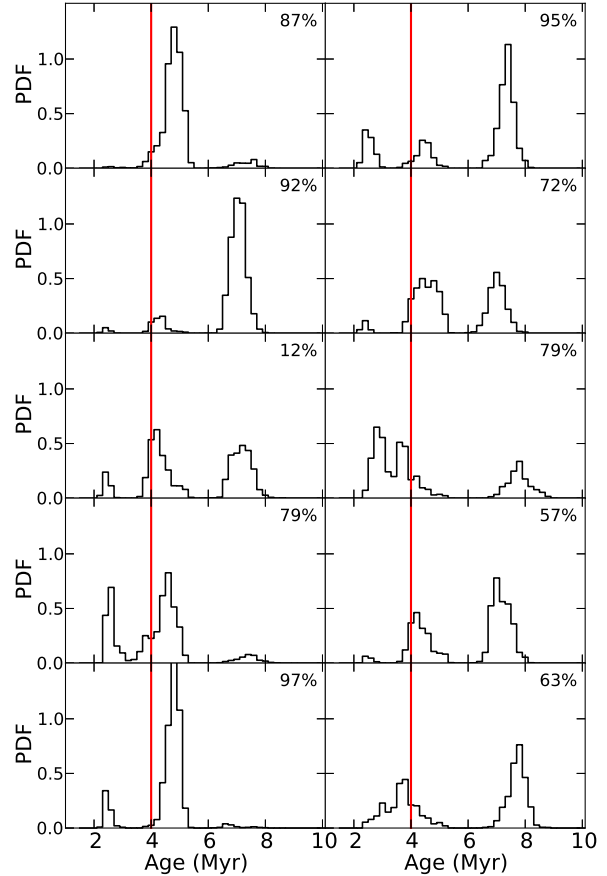}
\end{center}
\caption{
Test-clusters' age posterior probability distributions for 10 different
realizations of a cluster with $\alpha=1.7$, t$=$4 Myr, d$=$8 kpc,
M$_{cl,obs}=10^4$ \msun, and only single stars.
The input age of 4 Myr is marked ({\em vertical red}) and the
confidence interval that just includes 4 Myr is reported in
the top right corner of each panel.
The confidence interval is calculated by sorting the probability
distribution and integrating starting with the highest bins and
proceeding until the bin including 4 Myr is reached.
The input age always falls within the 99\% confidence interval.
If each peak is treated as an independent solution, the input age
is recovered with even greater confidence.
}
\label{fig:sim_age_monte_carlo}
\end{figure}

A cluster's age is often one of the most difficult parameters to constrain, especially
at young ages and when only high-mass main-sequence and post-main-sequence stars are
observed.
The number of WR stars and the ratios of WR to OB stars can be a precise indicator of
age; however, there are also cases where degeneracies produce several possible
solutions \citep{paumard06}.
To illustrate these degeneracies we simulate clusters at a range of ages from
1.5 - 10 Myr and examine the ratio of N$_{WR}$ to N$_{OB}$.
Note that N$_{OB}$ includes all OB stars down to Kp$=$15.5, which is equivalent
to B1$-$B2 V.
Figure \ref{fig:age_vs_WR_OB} shows that N$_{WR}$/N$_{OB}$ initially rises
around 2 Myr, peaks at 3 Myr, and falls to a minimum around 6 Myr, which is set almost
entirely by the rise and fall of N$_{WR}$.
After 6 Myr, the ratio rises slightly again as N$_{WR}$ remains relatively constant,
but N$_{OB}$ begins to drop.
If we observe a N$_{WR}$/N$_{OB}$ ratio of 0.125, then there are two possible
ages: 2.9 Myr and 3.8 Myr, assuming an IMF slope of $\alpha=1.7$.
This degeneracy is not entirely independent of the IMF slope
(Figure \ref{fig:age_vs_WR_OB}, right panel).
In order to explore how our bayesian methodology handles this age degeneracy,
we produce ten different realizations of a cluster, all with the same
input parameters (t$=$4 Myr, $\alpha=1.7$, M$_{cl,obs}=10^4$ \msun, d$=8$ kpc,
single stars).
Each cluster is then fit using the Bayesian inference methods described in
\S\ref{sec:bayesian_methodology}.
The resulting probability distributions for the age are shown in
Figure \ref{fig:sim_age_monte_carlo} and multiple age solutions are almost always found.
We note that in these 1D marginalized probability distributions, the input cluster age
is not always the highest peak.
However, the input age is always recovered within the 99\% confidence
interval and with much higher confidence if each peak is treated as an independent
solution in a multi-modal distribution.
Given the results of our testing, all further age results will be reported using
multi-modal solutions, if necessary.

\subsubsection{Multiplicity}

\begin{figure}
\begin{center}
\includegraphics[scale=0.5]{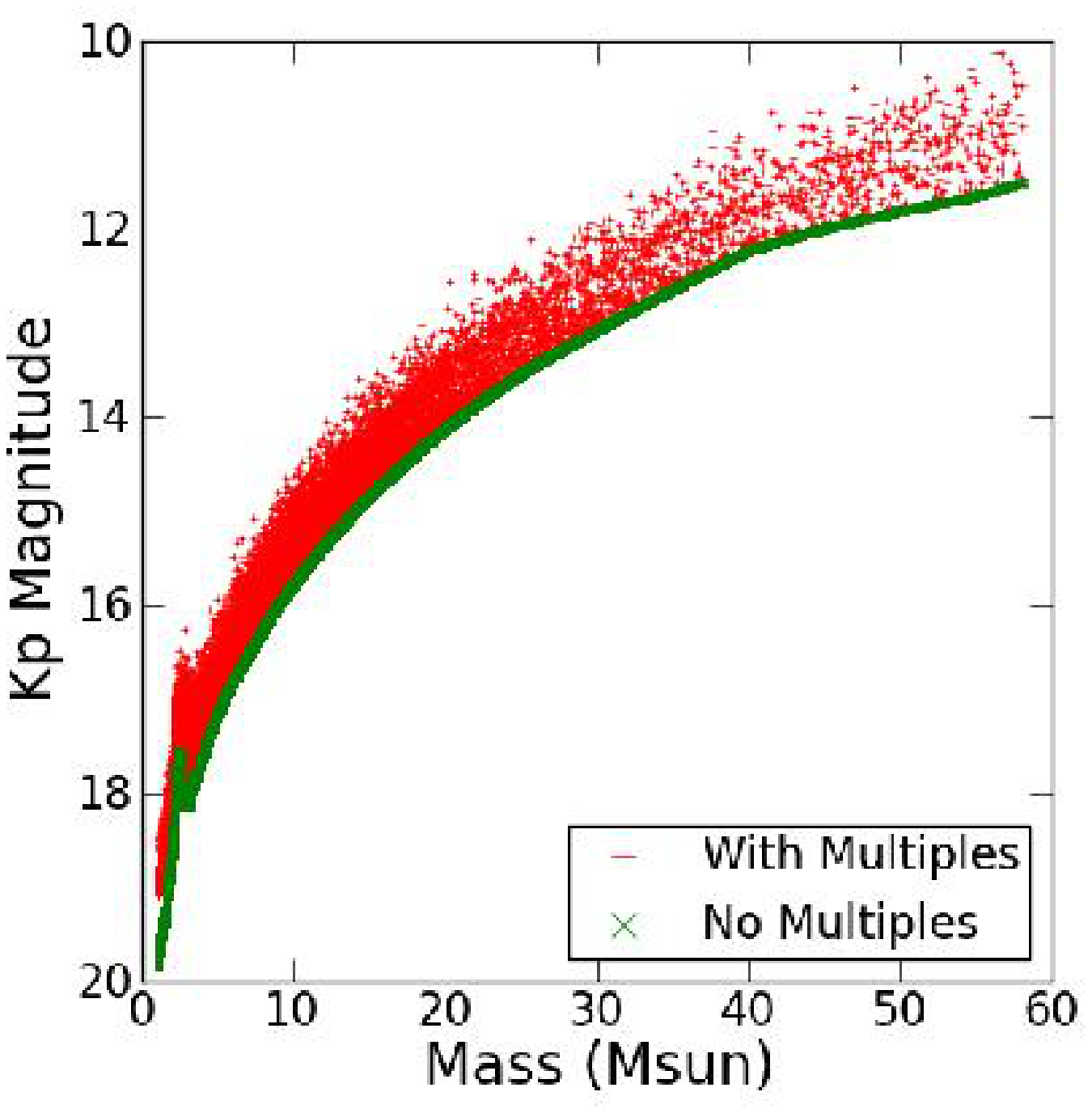}
\includegraphics[scale=0.5]{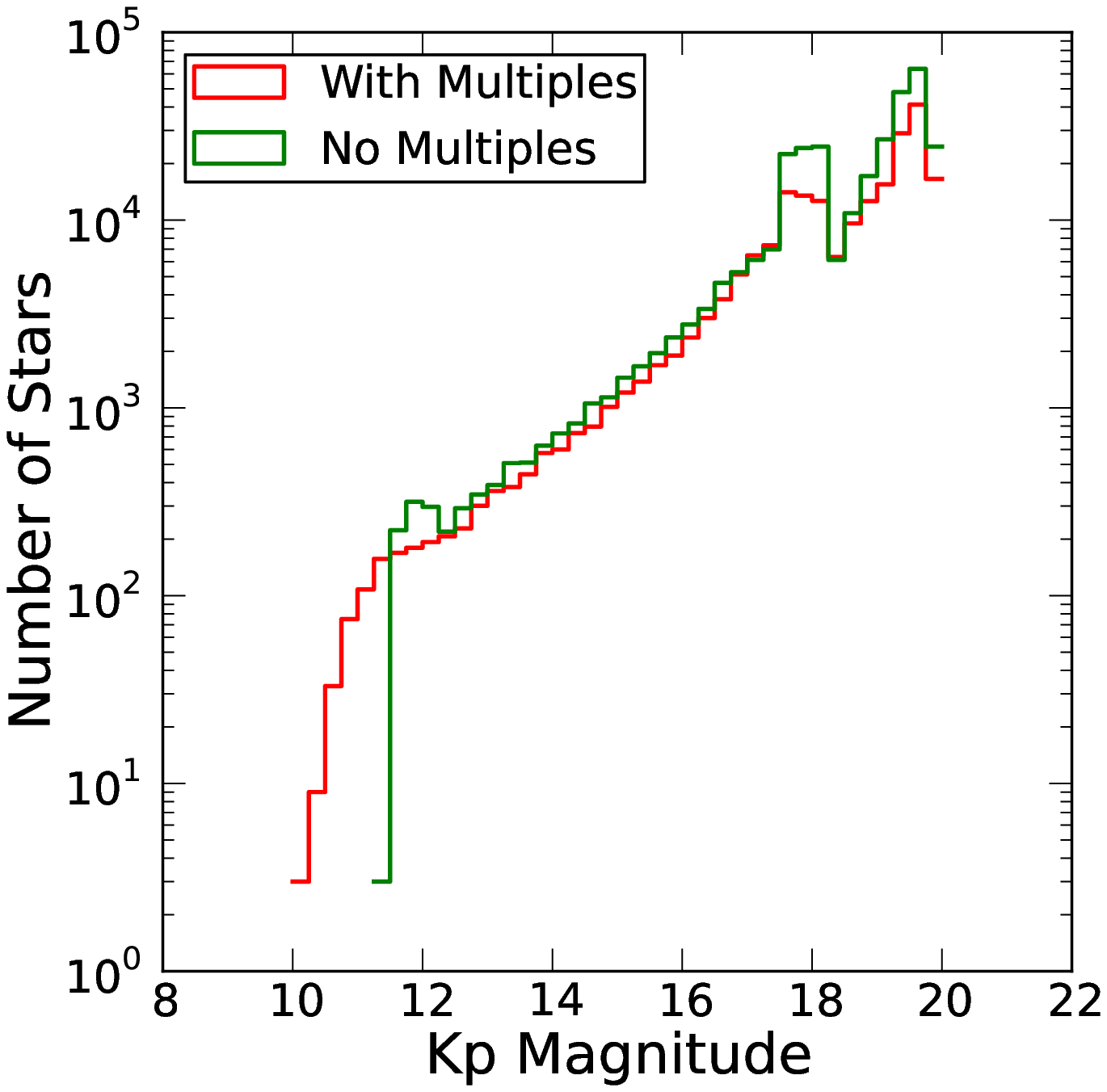}
\end{center}
\caption{
Comparison between the Kp distribution for model clusters with only single
stars ({\em green}) and  with multiple systems ({\em red}).
Both clusters have the same basic parameters:
$t=$4 Myr, $d=$8 kpc, $A_{Ks}=$2.7, and $M_{cl,y}=10^6$ \msun.
{\em Left:} The mass-magnitude relation for two simulated clusters
showing that multiple systems with high mass primaries are
typically 1 mag brighter than single stars.
{\em Right:} The binned Kp luminosity function for both clusters.
The cluster with multiple systems extends to higher Kp magnitudes than
does the cluster with only single stars. We also note the bump at
Kp$\sim$18 is the pre-main-sequence turn-on.
}
\label{fig:klf_vs_multiples}
\end{figure}

\begin{figure}
\begin{center}
\includegraphics[scale=0.5]{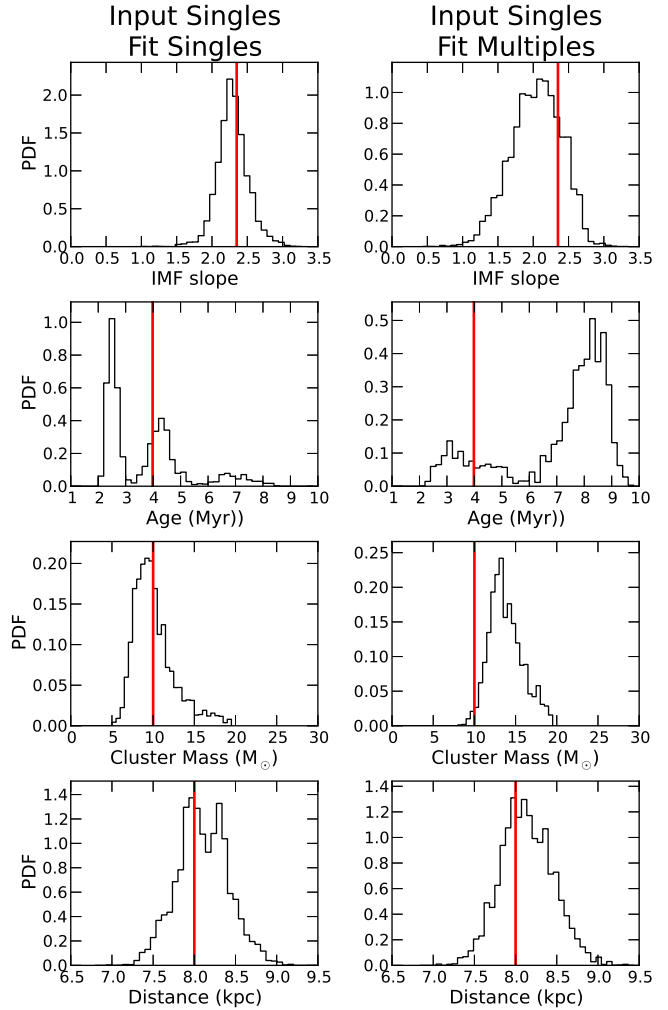}
\end{center}
\caption{
Test-cluster's marginalized 1D posterior probability density functions
(PDF) for a simulated cluster containing only single stars that is
subsequently fit allowing
only single stars {\em left} or multiple systems {\em right}.
The input values for the cluster's IMF slope, age, mass, and distance are
shown as a {\em vertical red line} in each panel.
For a cluster containing only single stars, fitting with multiple systems produces
extremely biased estimates for the cluster age and only slight biases in the IMF slope
and cluster mass. Without prior information, it is difficult to discern that
an inaccurate multiplicity model is being used.
}
\label{fig:sims_vs_multi_simS}
\end{figure}

\begin{figure}
\begin{center}
\includegraphics[scale=0.5]{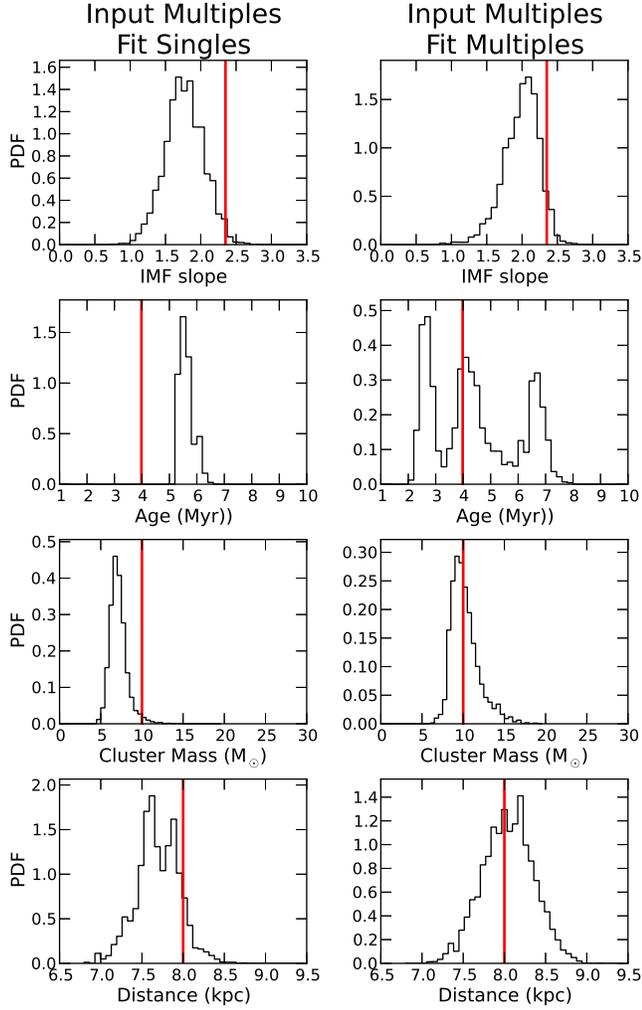}
\end{center}
\caption{
Test-cluster's marginalized 1D posterior probability density functions (PDF) for a simulated
cluster containing multiple star systems that is subsequently fit allowing
only single stars {\em left} or multiple systems {\em right}.
The input values for the cluster's IMF slope, age, mass, and distance are
shown as a red line in each panel.
For a cluster containing multiple systems, fitting with single stars produces
extremely biased estimates for nearly all the parameters.
Without prior information, it is easy to discern that
an inaccurate multiplicity model is being used since the distance tends toward
unphysically low values.
Multiple systems in clusters produce bright systems that are brighter
than is allowed in the model clusters with only single stars at 8 kpc.
Thus the only way to produce such bright systems is to move the cluster closer
in distance.
}
\label{fig:sims_vs_multi_simM}
\end{figure}

The need to incorporate multiplicity when determining a cluster's initial mass function
is often discussed \citep{kroupa1995,goodwin2005,weidner2009} but rarely implemented due
to incomplete knowledge of
multiplicity fractions, number of companions, companion mass functions, and how
these all scale with primary mass (see Appendix \ref{sec:multiplicity}
and references therein). The presence of multiple systems influences
the bright end of the KLF as illustrated in Figure
\ref{fig:klf_vs_multiples} by populating the brightest magnitude bins
(Kp$<$11) with sources that would otherwise be much fainter single stars.
We tested two possible scenarios to explore the impact of multiplicity in our analysis.
First, a cluster was simulated with only single stars and analysis
was done with and without allowing multiple systems in the fit.
Second, a cluster was simulated with multiple systems and, again, analysis
was done with and without allowing multiple systems in the fit.
Both clusters had an IMF slope of $\alpha=2.35$, an age of 4 Myr,
and a mass of $10^4$ \msun, although similar tests were performed with different
IMF slope and age combinations and the results are robust.
Figure \ref{fig:sims_vs_multi_simS} shows the results for the simulated
cluster containing only single stars.
Incorrectly fitting this cluster with multiple systems produces
extremely biased estimates for the cluster age and only slight biases in the IMF slope
and cluster mass. Without prior information, it is difficult to discern that
an inaccurate multiplicity model is being used.
Figure \ref{fig:sims_vs_multi_simM} shows the results for the simulated
cluster containing multiple systems.
Incorrectly fitting this cluster with single stars produces
extremely biased estimates for nearly all the parameters.
Even without prior information, it is easy to discern that
the single-star model is inaccurate since the distance tends toward
unphysical low values.
Multiple systems in clusters produce Kp magnitudes that are brighter
than is allowed in the model clusters with only single stars at 8 kpc.
Thus the only way to produce such bright systems is to move the cluster closer
in distance.
Shifting the cluster to younger ages is also a possibility; but such young clusters
do not give the correct number of WR stars.
The largest error is made when using single-star model clusters to fit
to clusters with multiple systems.
Thus, we choose to fit the observed data with simulated clusters containing
multiple systems.
All additional tests described below used simulated clusters containing multiple
systems and were fit with multiples.

\subsubsection{IMF Slope}

\begin{figure}
\begin{center}
\includegraphics[scale=0.3]{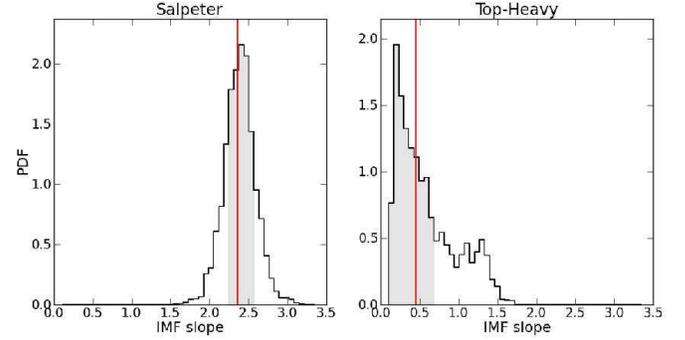}
\end{center}
\caption{
Test-cluster's marginalized 1D posterior probability density functions (PDF) for a simulated
cluster with IMF slope = 2.35 ({\em left}) and IMF slope = 0.45 ({\em right}).
The input values for the cluster's IMF slope, age, mass, and distance are
shown as a {\em vertical red} lines in each panel.
The grey shaded regions represent the 68\% confidence interval for the output
probability distributions and all of the input values fall within this
interval.
The confidence intervals are calculated by first finding the peak in
the probability distribution and then stepping away from the peak
until the integrated probability reaches 68\%.
}
\label{fig:sims_vs_imf_slope}
\end{figure}

Our bayesian inference methodology shows no systematic biases with respect
to the IMF slope as shown in the following tests on synthetic clusters.
Two clusters were simulated with an age of 6 Myr and with IMF slopes of
$\alpha=2.35$ (Salpeter) and $\alpha=0.45$ (top-heavy reported by \citet{bartko2010}).
The Salpeter-like cluster has an input mass of 10,000 \msun and the top-heavy
cluster has an input mass of 40,000 \msun in order to produce $\sim$100 non-WR stars
with Kp$\leq$15.5; similar to our observed data set.
Figure \ref{fig:sims_vs_imf_slope} shows the output posterior probability distributions
for the IMF slope, age, and cluster mass for the two simulated clusters.
The input IMF slope falls well within the 68\% confidence interval of the posterior
probability density function.
A handful of similar cluster tests were performed with different ages and masses
and the input and output IMF slopes always agree within the 68\% confidence region and
there appears to be no significant bias to either higher or lower IMF slopes.

\section{Results}
\label{sec:results}

\begin{figure}
\begin{center}
\includegraphics[scale=0.4]{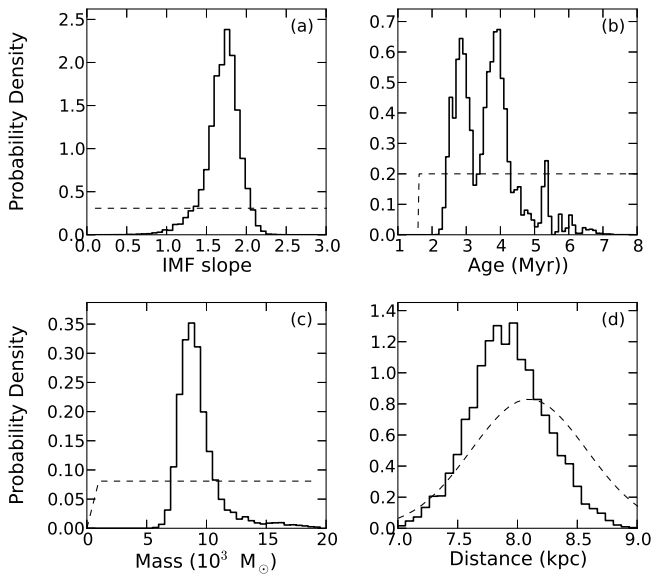}
\end{center}
\caption{
Observed YNC's marginalized 1D posterior probability density functions (PDF) for the cluster properties.
The {\em histograms} show results from the Multinest bayesian analysis including multiple systems.
The {\em dashed black} lines show the prior probability distributions used.
The resulting constraints on the IMF slope ({\em panel a}), the cluster age ({\em panel b}) and
the cluster mass ({\em panel c}) are significant compared with the prior PDFs used.
The distance constraint ({\em panel d}) is weak and is largely a reflection of the prior.
}
\label{fig:fit_pdf_1d}
\end{figure}

\begin{figure}
\begin{center}
\includegraphics[scale=0.4]{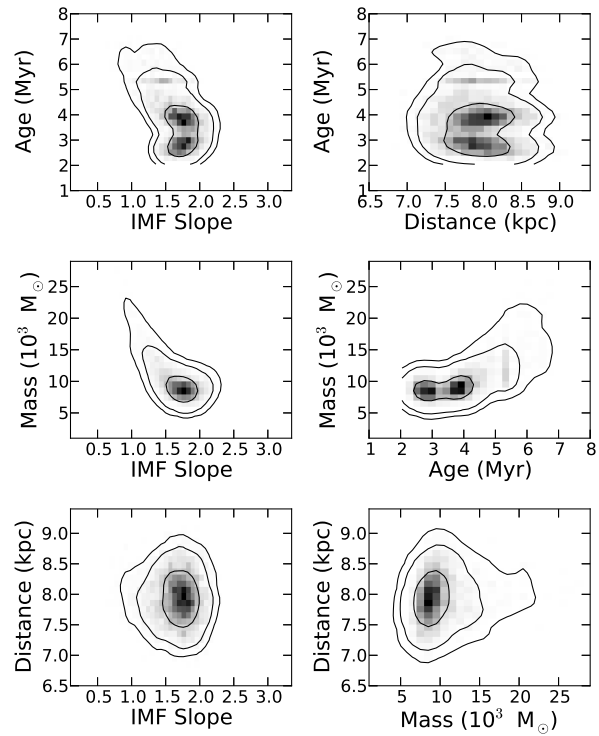}
\end{center}
\caption{
2D posterior probability density functions (PDF) for the observed cluster's properties.
The over-plotted contours give 68\%, 95\%, and 99\% confidence intervals.
Weak correlations exist between age, mass, and IMF slope.
The correlation between the cluster mass and the age or
IMF slope is a consequence of the age-IMF slope relationship since,
at older ages, the most massive stars have dissapeared and the
cluster mass must increase to match the observed numbers of stars
brighter than Kp$=15.5$.
}
\label{fig:fit_pdf_2d}
\end{figure}

\begin{figure*}
\begin{center}
\includegraphics[scale=0.5]{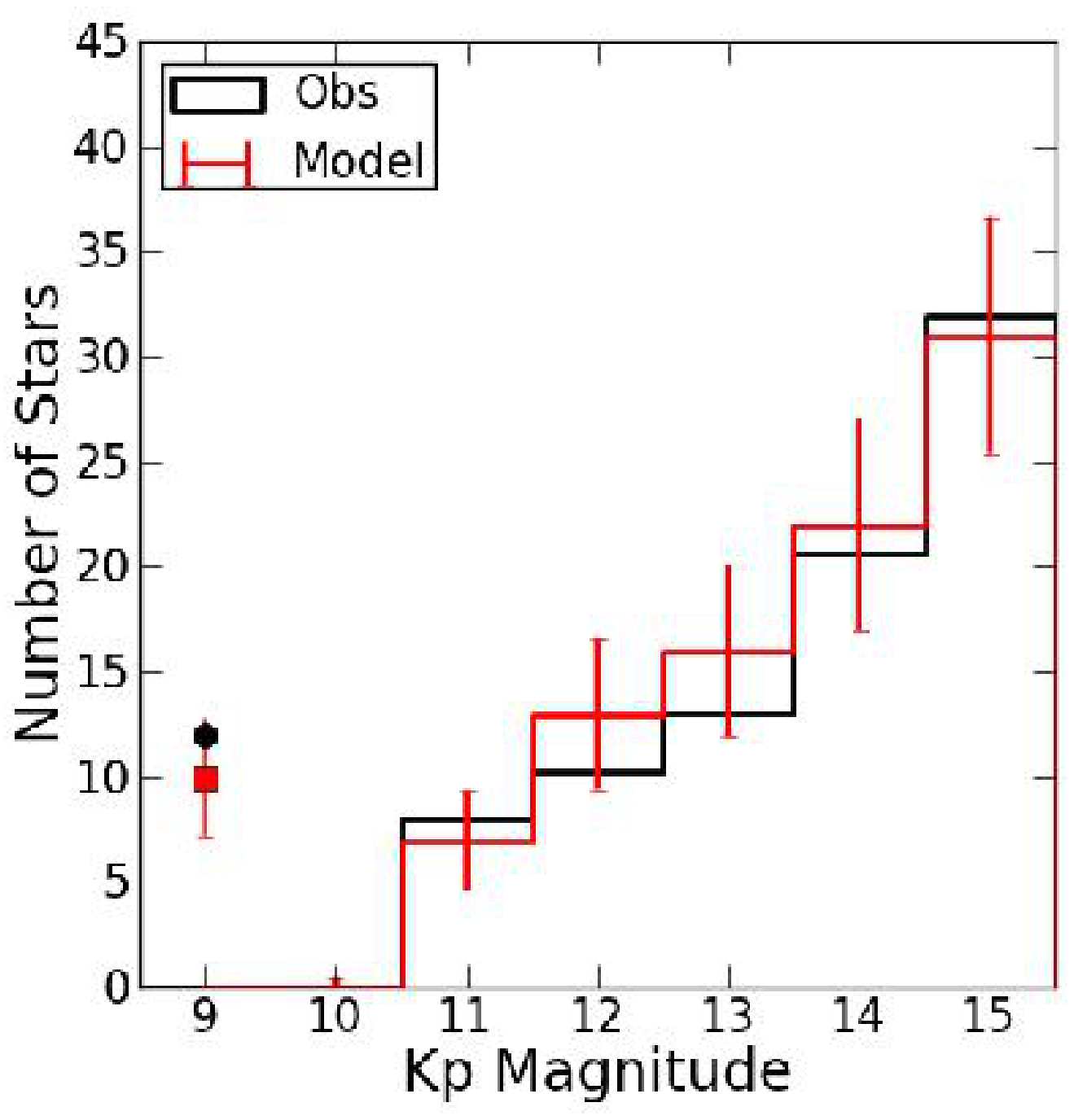}
\includegraphics[scale=0.5]{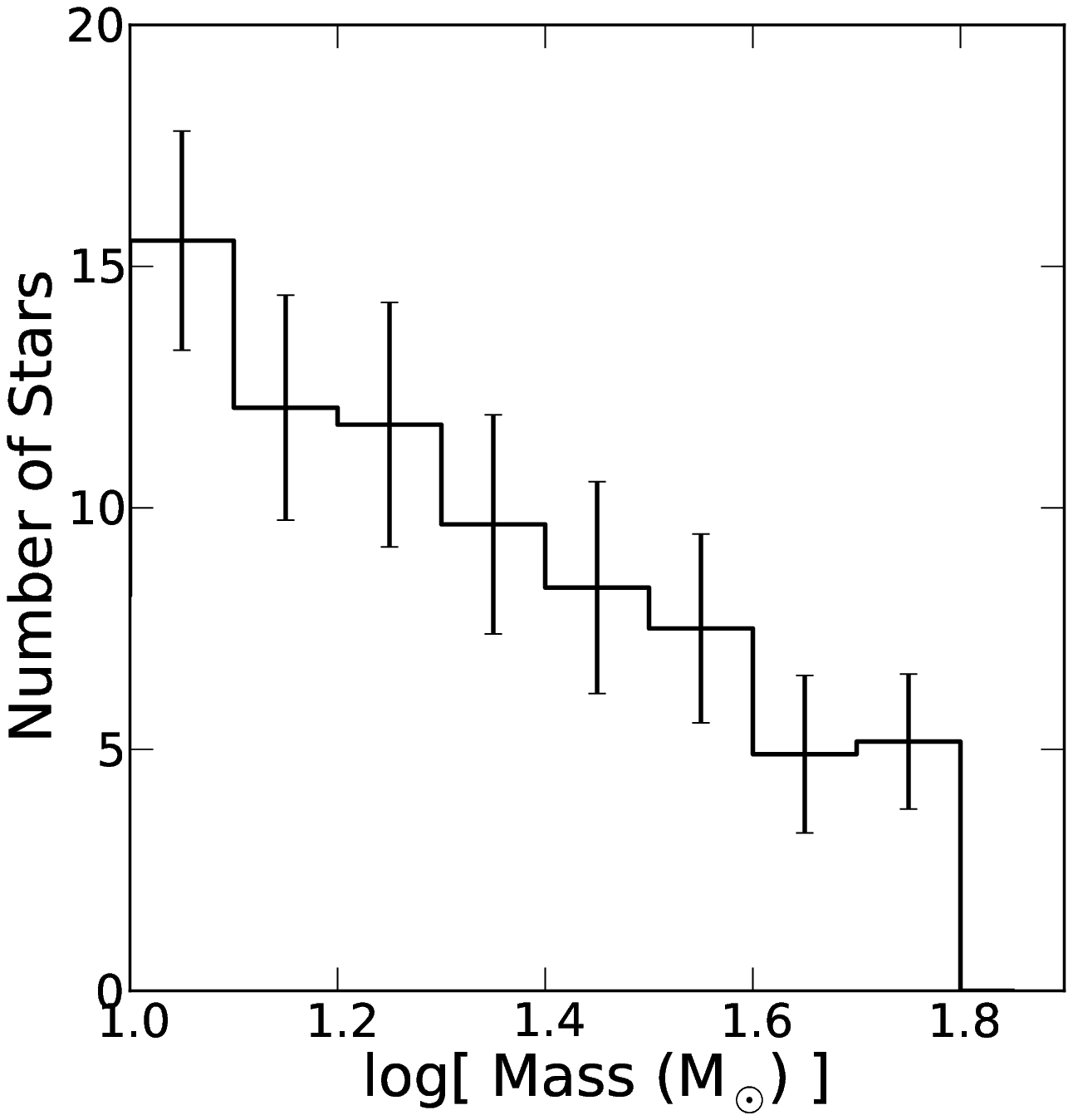}
\end{center}
\caption{
The observed ({\em black}) and model ({\em red}) Kp-band
luminosity function ({\em left}) and the resulting mass function ({\em right}).
The model cluster includes multiple systems and has the
properties listed in Table \ref{tab:cluster_properties} (Solution 1).
({\em Left}):
The observed KLF ({\em black line}) and number of WR stars ({\em black circle}) show
good agreement with the model KLF ({\em red line}) and model WR stars
({\em red square}).
The observed Kp magnitudes ({\em red}) are corrected for differential extinction to a common extinction value of A$_{Ks}=2.7$.
The model KLF is taken as the mean and standard deviation ({\em red errorbars})
of 100 simulated clusters, all with the same parameters, in order to account for
variations due to stochastic sampling of the IMF.
The imaging completeness curve is applied to each simulated cluster for
direct comparison to the observed KLF.
({\em Right}):
The resulting mass function is constructed by
converting the observed stars' Kp magnitudes to masses using the
best-fit isochrone.
To capture uncertainties in the stars' brightness,
each star's Kp magnitude is randomly sampled 100 times from
a Gaussian distribution centered on the measured Kp with a width of $\sigma_{Kp}$.
The resulting masses are binned into a mass function and
the mean and standard deviation of the 100 mass functions are shown as
a {\em black line} with errorbars.
The Wolf-Rayet stars are not included in the mass function.
}
\label{fig:fit_KLF_and_WR_stars}
\end{figure*}

\begin{figure}
\begin{center}
\includegraphics[scale=0.35]{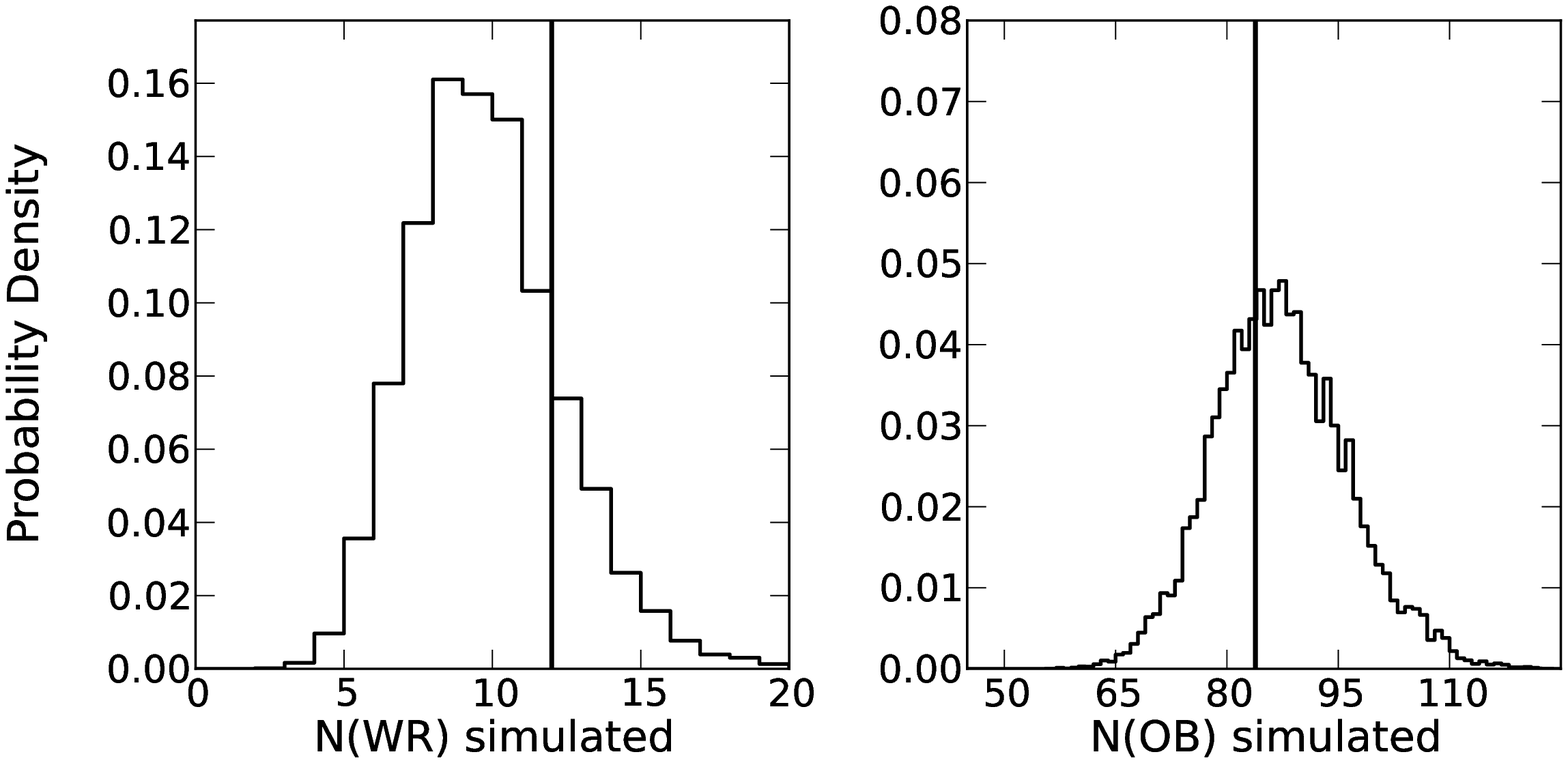}
\end{center}
\caption{
Comparison of the observed vs. inferred distributions for the number of Wolf-Rayet stars ({\em left}) and the number of OB stars ({\em right}). The observed $N_{WR}$ and $N_{OB}$ are shown as {\em vertical black lines} and the probability distributions from the Bayesian analysis are shown as histograms.
}
\label{fig:fit_N_pdf_1d}
\end{figure}

The observed Kp magnitudes, their uncertainties, and the number of Wolf-Rayet and OB stars
were used in Bayesian inference to determine the cluster's physical properties.
Figure \ref{fig:fit_pdf_1d} shows the resulting 1D posterior probability distributions
for the 4 free parameters in the model (cluster age, mass, distance, and IMF slope).
The model assumes that multiple systems are present with the properties described in
\S\ref{sec:synthetic_clusters} and Appendix \ref{sec:multiplicity}.
Several parameters show correlations including a moderate correlation
between the cluster age and IMF slope (Figure \ref{fig:fit_pdf_2d}).
The correlation between the cluster mass and the age or
IMF slope is a consequence of the age-IMF slope relationship since,
at older ages, the most massive stars have dissapeared and the
cluster mass must increase to match the observed numbers of stars
brighter than Kp$=15.5$.
The posterior probability distributions are the most accurate representation
of the results; however, we also present ``best-fit'' values in
Table \ref{tab:cluster_properties} represented by the expectation value and
68\% and 95\% Bayesian confidence intervals\footnote{Central credible intervals.}
of the marginalized 1D posterior probability density functions.
The resulting age distribution is multi-modal.
Therefore, we report solutions from three possible samples:
(1) ages greater than 3.3 Myr,
(2) ages less than or equal to 3.3 Myr, and
(3) the complete posteriors,
The age boundaries for the two modal solutions were estimated
based on the minima between the two peaks.
We report all parameters for these three solutions; however,
all solutions have similar $\alpha$, M$_{cl,obs}$, and $d$
distributions.
We will adopt Solution 1 in the remainder of the paper as it contains
the most probable solution (with the maximum likelihood) and contains the bulk
of the probability density.
Solution 1 is also favored over the smaller age in Solution 2 based
on more detailed spectroscopic analyses that show the Wolf-Rayet star
sub-types and the supergiant nature of the brightest OB stars favor
ages $\gtrsim$4 Myr \citep{paumard06,martins07}.
A comparison of the observed and model KLFs and mass functions from the
inferred parameters of Solution 1 is shown in Figure \ref{fig:fit_KLF_and_WR_stars}.
The complete posteriors for the number of predicted Wolf-Rayet and OB
stars are also shown in
Figure \ref{fig:fit_N_pdf_1d} and are compared with the observed number of WR and OB stars.
The overall fit is good given the theoretical uncertainty in the evolution of
massive stars and the unknown mapping between progenitor mass and Wolf-Rayet phases.

\begin{deluxetable*}{lccc}
\tablewidth{0pt}
\tablecaption{Cluster Properties and Uncertainties}
\tablehead{
Cluster & Expectation & 68\% & 95\% \\
Properties & Value & Interval & Interval
}
\startdata
Solution 1: Age$>$3.3 Myr               & & & \\
$\;\;\;$Age (Myr)                       & 4.2  &  3.6 - 4.8  & 3.4 - 6.1  \\
$\;\;\;$IMF slope ($\alpha$)            & 1.7  &  1.5 - 1.9  & 1.1 - 2.1  \\
$\;\;\;$Integrated Mass ($10^3$ \msun)    & 10.1 &  8.1 - 11.6 & 7.2 - 18.5 \\
$\;\;\;$Distance (kpc)                  & 7.9  &  7.6 - 8.3  & 7.3 - 8.6  \\
Solution 2: Age$\leq$3.3 Myr            & & & \\
$\;\;\;$Age (Myr)                       & 2.8  & 2.6 - 3.1  & 2.4 - 3.3  \\
$\;\;\;$IMF slope ($\alpha$)            & 1.8 &  1.6 - 1.9  & 1.5 - 2.1  \\
$\;\;\;$Integrated Mass ($10^3$ \msun)    & 8.8 &  7.7 - 9.5  & 6.8 - 10.5 \\
$\;\;\;$Distance (kpc)                  & 7.9 &  7.6 - 8.2  & 7.3 - 8.5  \\
Total                                   & & & \\
$\;\;\;$Age                             & 3.6  & 2.8 - 4.3  & 2.5 - 5.8  \\
$\;\;\;$IMF slope ($\alpha$)            & 1.7 &  1.5 - 1.9  & 1.2 - 2.1  \\
$\;\;\;$Integrated Mass ($10^3$ \msun)    & 9.5 &  7.9 - 10.6 & 7.1 - 16.7 \\
$\;\;\;$Distance (kpc)                  & 7.9 &  7.6 - 8.3  & 7.3 - 8.6  \\
\enddata
\tablecomments{The integrated masses correspond to that within the
field of view of the survey (see Paper I for details) between
1$-$150 \msun.
The estimated mass for the entire YNC is 2-4 times larger. }
\label{tab:cluster_properties}
\end{deluxetable*}

\begin{figure}
\begin{center}
\includegraphics[scale=0.35]{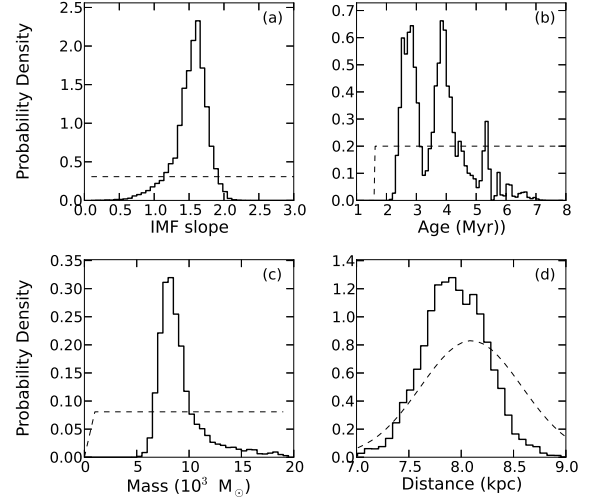}
\end{center}
\caption{
Same as Figure \ref{fig:fit_pdf_1d}, but excluding stars with
r$<$0\farcs8 to better match the sample chosen in \citep{bartko2010}.
The peak of the probability distribution for the IMF slope shifts to $\alpha=$1.6;
however, it is consistent with the result from the complete sample, within the
68\% uncertainties. IMF slopes as top-heavy as $\alpha=0.45$ remain inconsistent
at 99.89\% probability. Ages of 6-7 Myr become slightly more probable.
}
\label{fig:fit_pdf_1d_r0.8}
\end{figure}

For the IMF slope, recall that a flat prior from 0.5$-$3.5 was adopted; but the
posterior distribution is strongly peaked at 1.7 $\pm$ 0.2.
The slope is flatter than a typical Salpeter slope and slopes of 2.35 or
greater are ruled out at 3.8$\sigma$ (99.98\%).
The slope is inconsistent with the previously reported top-heavy slope of
0.45 \citep{bartko2010} at the 3.8$\sigma$ level (99.98\%).
Our sample of young stars includes those within r$<$0\farcs8; however, the results
are very similar even when these stars are excluded (Figure \ref{fig:fit_pdf_1d_r0.8},
as is done in \citet{bartko2010}.

The cluster mass is 10,100 \msun (68\% interval: 8,100$-$11,600 \msun)
between 1$-$150 \msun and is well
constrained by the data, given that the posterior distribution is significantly
different from the prior.
The cluster age is also significantly different from the prior and shows multiple peaks.
The most probable solution falls at 3.9 Myr (68\% interval: 3.6$-$4.8 Myr);
although the second peak at 2.8 Myr is still highly probable.
This multi-modal behavior is a result of the rapid changes in the number of WR
stars and the ratio of WR to OB stars in very short periods of time
(Figure \ref{fig:age_vs_WR_OB}).
Although the distance is also a free parameter, the resulting probability distribution
mainly reflects the prior shown in the dashed line.
This is reassuring since, in our testing, incorrect model assumptions (e.g. fitting single
star models to data with multiple systems) often led to skewed distance distributions,
which we do not observe using the multiple systems model.

The cluster mass and the numbers of WR and OB stars are based only on the
young stellar population within the field of view observed for this
study, which has incomplete azimuthal coverage.
This does not represent the entire cluster mass of the young nuclear star cluster.
The total cluster mass can be estimated by extrapolating beyond the
field of view in this study, in several different ways.
First, we use the assumption that the cluster is
spherically symmetric and has a surface density profile that falls as
$R^{-\Gamma}$ with $\Gamma = 0.9$ \citepalias{do2012}.
This yields a surface density weighted coverage of 27\% and a coverage-corrected
total cluster mass of $\sim$37,000 \msun within a projected radius
of 13\farcs65 and above 1 \msun.
The above assumption fails to account for known kinematic structures in
the young population, including a highly inclined disk containing as much as 50\%
of the young stars.
Furthermore, the above mass estimate assumes a constant IMF at all radii,
which is unlikely given the mass segregation that will occur over 4 Myr.
As an alternative, we instead make an empirically based estimate of the
completeness by taking the ratio of the number of young stars with Kp $<$ 13
within our FOV to the same number reported in \citet{paumard06}, which
we assume is 100\% complete out to 14''.
This yields a coverage completeness of 71\% and a total cluster mass of
$\sim$14,000 \msun above 1 \msun.
Thus, we estimate the total cluster mass to be in the range of 14,000 - 37,000 \msun
above 1 \msun.
We note that the total cluster mass depends on the low-mass cutoff as
M$_{cl} \cdot$ (m$_{lo}$ / 1 \msun)$^{-1.7}$; and, while
we have assumed 1 \msun, the lowest-mass young stars we observe are $\sim$8 \msun.
In order to improve the constraints on the total cluster mass, more sensitive and
azimuthally complete spectroscopic surveys at large radii are necessary.

\begin{figure}
\begin{center}
\includegraphics[scale=0.35]{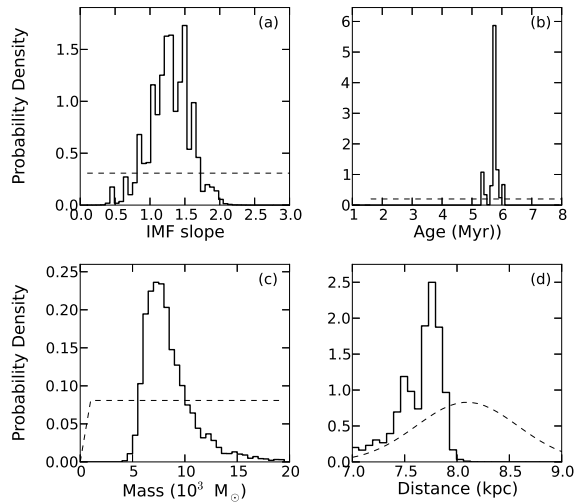}
\end{center}
\caption{
Same as Figure \ref{fig:fit_pdf_1d}, but only allowing single stars in
the model.
The overall fit of the single-star model is poor and produces a strong
bias away from the measured distance of 8 kpc.
Furthermore, the resulting evidence for this model is substantially
lower than for the model that includes multiples.
}
\label{fig:fit_pdf_1d_single}
\end{figure}

The above results are derived from modeling clusters with multiple
systems; however, for completeness, we also show the resulting posterior
probability distributions for a single-star model in
Figure \ref{fig:fit_pdf_1d_single}.
It is clear that the fit is poor.
The bias to very small distances indicates that the brightest OB stars combined
with the number of WR stars cannot be accurately fit with a single star model.
Our modeling indicates that multiple star systems are necessary to fit
the observed stellar population.
However, we caution that the multiplicity properties
were fixed to solar neighborhood values that may not be applicable to the
Young Nuclear Cluster.
Both the star formation process and subsequent dynamical processing may impact
multiplicity.

\section{Discussion}
\label{sec:discussion}
The young nuclear star cluster at the Galactic center, at least along
the plane of the well-defined clockwise disk, has an
IMF slope in the range $\alpha=1.5-1.9$ (68\% confidence interval) for stellar masses
above $\sim$10 \msun.
This is significantly flatter than clusters in the Milky Way disk and field ($\alpha=2.35$).
Our measured IMF is in good agreement with X-ray observations showing that
X-ray emission from pre-main-sequence stars is reduced by a factor of 10
compared with what is expected from observations of the Orion Nebular Cluster,
scaled to the age and distance of the Galactic center \citep{nayakshinSunyaev06}.
Extrapolating our measured IMF slope down to lower masses produces a factor
of 4$-$15 fewer X-ray emitting stars ($0.5-3.0$ \msun) than would be expected for
a slope of $\alpha=2.35$.

An extremely top-heavy IMF is inconsistent with our observations at high significance,
in contradiction to previous observations \citep{bartko2010}.
An extensive discussion of the possible observational and analysis
differences between our work and previous work is presented in \citetalias{do2012}.
Here we reiterate a few key differences.
The \citet{bartko2010} field of view extends predominantly
North, out of the plane of the well-defined clockwise disk of young stars.
They also remove all young stars in the central 0\farcs8 and outside 12''.
In the work presented here, the spectroscopic window in which young stars are identified
extends from the center to the ESE along the well defined clockwise disk structure.
And the entire population of young stars, including those within 0\farcs8
of the SMBH, is included in the analysis.
We assume that all the young stars were formed in a
single star formation event with the same physical conditions giving rise
to a constant IMF for all the kinematic sub-groups.
As discussed in Sections \ref{sec:intro} and \ref{sec:obs_and_analysis},
the stars in this innermost region were dynamically injected and it still appears
to be theoretically possible that they originated from the most recent starburst
that produced the young nuclear cluster.
Furthermore, restricting our sample of young stars to a radius
range of 0\farcs8$-$12'' does not significantly change the KLF shape.
Repeating our Bayesian analysis on this restricted sample produces a similar
posterior probability distribution for the IMF slopes to that of our full sample.

We find a cluster that is well fit with an instantaneous star formation event at an
age of 2.5-5.8 Myr with 95\% confidence.
The most probable age is 3.9 Myr.
This is smaller than the 6 Myr commonly adopted,
although consistent within the uncertainty range reported in
\citet{paumard06}.
The age of the cluster is mainly constrained by the number of WR stars
and the highest luminosity OB stars detected, since this effectively gives a
main-sequence turn-off age.
Careful examination of Figure 11 in \citet{paumard06} shows that best-fit ages,
given the different numbers of O and WR stars (WN, WC, WNE, WNL), are actually
either 4 Myr or 8 Myr, with 6 Myr being improbable for a Salpeter IMF.
The inclusion of fainter OB stars in our spectroscopic sample
substantially reduces the probability of cluster ages as old as 8 Myr.
Figure \ref{fig:age_vs_WR_OB} shows the expected number of WR and OB stars
and their ratio as a function of cluster age for different assumed
mass functions.
While our Bayesian analysis captures uncertainties in the age due to random
sampling of the mass function, it does not capture systematic uncertainties in
models of stellar evolution.
The lifetimes and luminosities of WR stars are sensitive to metallicity
(Fe and $\alpha$ elements), mass-loss rates, initial rotation rates, and close-binary
mass exchange.
Increased metallicity, mass-loss rates, rotation rates, and binary mass exchange
all tend to increase the ages at which WR stars still exist
\citep{eldridgeVink2006,meynet2006}.
On the other hand, increased metallicity reduces the main-sequence lifetimes.
Furthermore, massive clusters, such as Westerlund 1, contain WR stars, blue supergiants,
yellow hypergiants, and red supergiants at an age of 5 Myr; which no stellar evolution
model successfully predicts \citep[e.g.][]{negueruela2010}.
Therefore, we caution that ages up to 8 Myr may still be reasonable given current
uncertainties in massive-star evolutionary models.
We note that the IMF slope measurement is dominated by the relative luminosity
distribution of OB main-sequence and supergaint stars,
rather than the numbers of windy WR stars, and is far
less sensitive to systematic uncertainties in evolutionary models.
If the young nuclear cluster does have a slightly younger age of $\sim$4 Myr,
then there are important implications for the dynamical history
of the cluster. For instance, models of {\em in situ} star formation with a circular
gas disk as the initial condition require at least 6 Myr for the
orbital eccentricities to evolve from circular to the observed
eccentricity distribution peaked at $e=0.3$, assuming a Salpeter
IMF \citep{alexander07imf,cuadra08}.
Revised calculations are necessary to determine whether the
observed IMF slope and younger age can give rise to the observed high
eccentricities in such a short time.

\begin{figure}
\begin{center}
\includegraphics[scale=0.5]{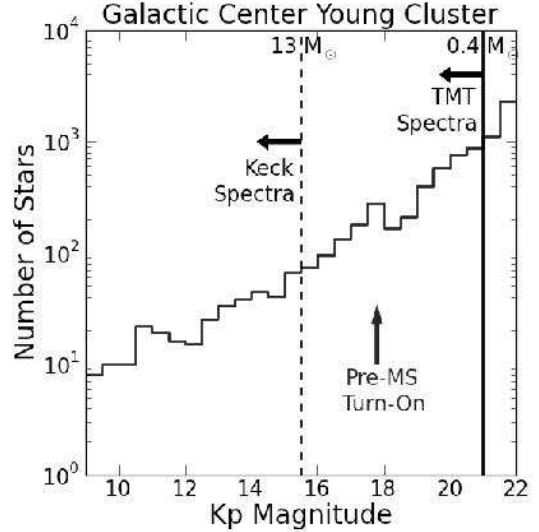}
\end{center}
\caption{
A model Kp-band luminosity function for the Young Nuclear Cluster showing the
relative sensitivity of Keck spectroscopy (Kp$<$15.5) and future TMT+IRIS
spectroscopy (Kp$<$21).
With TMT+IRIS, the pre-main-sequence turn-on point will be detectable at Kp$\sim$17.5
as well as the full IMF shape down to $\sim$0.4 \msun.
The model cluster shown here has an IMF slope of 1.7 down to 0.5 \msun
and then turns over at 0.5 \msun to a slope of 1.3 down to 0.1 \msun as
suggested by \citet{weidnerKroupa2004} for star formation in the local neighborhood.
The model cluster has a distance of 8 kpc and an extinction of A$_{Ks}=2.7$.
}
\label{fig:klf_with_tmt}
\end{figure}

Ultimately, a complete spectroscopic and astrometric census of the young population in the
Young Nuclear Cluster is required to accurately measure the recent star formation history
and determine if the distinct kinematic components were formed under different
conditions, resulting in different mass functions.
Expanding the survey area with current facilities will help clarify whether young stars
on and off the disk have a different IMF.
The Bayesian methodology presented here should be expanded to
incorporate additional information from spectroscopy, such as effective temperature
and gravity constraints for the OB stars, surface-abundance sub-types for the Wolf-Rayet
stars, and luminosities/temperatures from Wolf-Rayet wind/atmosphere modeling
such as presented in \citet{martins07}.
This additional information may improve age constraints; although uncertainties
in stellar evolution for massive stars will still be a major factor.
In the future, detection of pre-main-sequence turn-on points at Kp=17.5 would
more definitively constrain the age and star formation history.
Furthermore, to truly constrain physical models of star formation at the Galactic center
relative to local star formation processes, we will need measurements of the IMF shape over
a broader mass range, ideally including any turnover suggesting a characteristic mass scale
(e.g. $\sim$0.5 \msun in Orion).
Some theories of star formation suggest that in extreme
conditions, such as those found at the Galactic center, the characteristic mass
may be significantly higher than in local star-forming regions
\citep{morris93,bonnell04,krumholzMckee08}.
Measurements of both the characteristic mass and the pre-main-sequence turn-on point
should be possible with the next generation of large
ground-based telescopes equipped with adaptive optics-fed integral field spectrographs.
One such instrument is IRIS on the Thirty Meter Telescope (TMT), which can
accurately spectral-type stars in the Galactic center down to Kp$=$21 in 1 hour
\citep{wright2010}, as shown in Figure \ref{fig:klf_with_tmt}.
Predicted IRIS sensitivities reach 0.4 \msun stars (Kp$=$21); however, distinguishing
the young low-mass objects from the sea of much older main-sequence stars may require
substantially higher signal-to-noise to make precise measurements of effective
temperatures.
Such observations will be essential as the Young Nuclear Cluster is one of the few
environments where significant deviations have been found from an otherwise
near-universal IMF.

\section{Conclusions}
The Galactic center hosts a young nuclear star cluster around the
supermassive black hole.
We use membership probabilities derived from spectroscopy,
precise infrared photometry, and Bayesian inference
methods to determine the global properties of the cluster. The
best-fit age of the cluster is 3.9 Myr, somewhat younger than
previously reported. The cluster's total mass extrapolated down to 1 \msun
is between 14,000 - 37,000 \msun after correcting for incomplete azimuthal
coverage. The best-fit IMF slope of 1.7 is flatter than a traditional
Salpeter IMF; but far steeper than previously claimed in the
literature. Future spectroscopic observations covering a larger field
of view and extending to lower masses are necessary to understand
the impact of mass segregation on IMF slope measurements and to
determine whether the peak in the initial mass function also
shows any significant difference from nearby star-forming regions.

\acknowledgements
We would like to thank Greg Martinez for suggesting the MultiNest program and
for helpful suggestions on our statistical analysis.
J.~R. Lu wishes to acknowledge support from the California Institute of Technology
Millikan Postdoctoral Fellow program and the NSF Astronomy and Astrophysics Postdoctoral
Fellow program (AST$-$1102791). We also acknowledge support from NSF grant
AST-0909218 (PI Ghez), and the Levine-Leichtman family foundation.
The data presented herein were obtained at the W.M. Keck Observatory, which is operated
as a scientific partnership among the California Institute of Technology, the University of
California and the National Aeronautics and Space Administration. The Observatory was made
possible by the generous financial support of the W.M. Keck Foundation. The authors wish to
recognize and acknowledge the very significant cultural role and reverence that the summit
of Mauna Kea has always had within the indigenous Hawaiian community.  We are most fortunate
to have the opportunity to conduct observations from this mountain.

\appendix

\section{Multiplicity}
\label{sec:multiplicity}

\begin{figure}
\begin{center}
\includegraphics[scale=0.5]{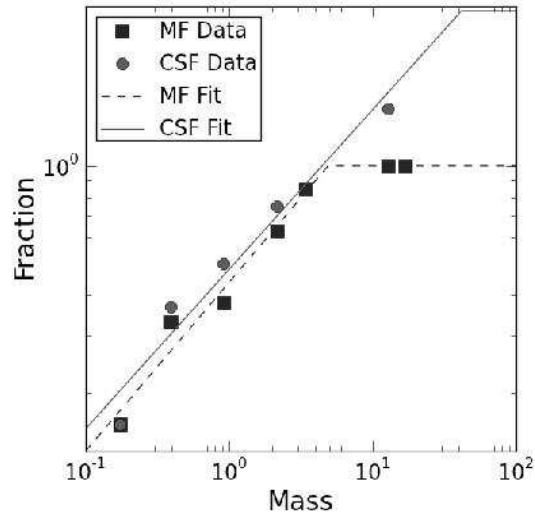}
\end{center}
\caption{
The multiplicity frequency (MF) and the companion star frequency (CSF) as a function
of primary mass.
Empirical measurements are plotted for the MF ({\em blue squares}) and
CSF ({\em red circles}) values reported in the literature (Table \ref{tab:multiplicity_literature}).
The data are fit with power laws ({\em lines}) as given by Equations \ref{eqn:multi_MF} and \ref{eqn:multi_CSF}.
The CSF is truncated at $\leq$3, which impacts systems with a primary mass
above $\sim$40 \msun.
}
\label{fig:binary_properties}
\end{figure}

The distribution of multiple star systems can be described completely by equations for the
multiplicity frequency (MF), the companion star frequency (CSF), the mass ratio (q), and the
separation distribution.
For star clusters at the distance of the Galactic center (8 kpc), multiple systems
are spatially unresolved and the separation distribution can be integrated over.
The multiplicity frequency and companion star frequency are defined as in
\citet{reipurthZinnecker1993}
\begin{eqnarray}
MF = \frac{B + T + Q + …}{S + B + T + Q + ...} \\
CSF = \frac{B + 2T + 3Q + …}{S + B + T + Q + ...}
\end{eqnarray}
where S is the number of single stars, B is the number of binaries,
T is the number of triples, and Q is the number of quadruples.
The multiplicity fraction always ranges between 0 and 1; but the companion star fraction,
which is the mean number of companions, can be greater than 1, such as for the Orion
Trapezium stars where CSF$>$1.5 \citep{preibisch1999,zinneckerYorke2007}.
Based on observations, the MF and CSF are known to vary with primary star mass,
the age of a cluster, and possibly the density of a cluster
\citep[e.g.][]{ghez1993,petr1998,mason1998,duchene1999,kohler2000,reipurth2000,patience2002,shatsky2002,kohler2006,duchene2007,kohler2008,mason2009,raghavan2010,kraus2012}.
Complete multiplicity surveys of {\em young} clusters ($<$10 Myr) that span a large
range of primary stellar masses (0.1 $-$ $>10$ \msun), mass ratios,
and separations are difficult to conduct and only a few exist in the literature
\citep{lafreniere2008,kobulnickyFryer2007,kouwenhoven2007}.
Star formation theories do not yet predict an analytic functional form for the MF, CSF,
and q distributions and how they vary with primary mass; however, simulations are
beginning to produce distributions that are in rough agreement with observations
\citep{bate2012,krumholz2012}.
Therefore, we take an empirical approach and compile measurements of the MF and CSF
as a function of mass from published surveys of young clusters ($<$10 Myr), which span
stellar masses of 0.2$-$17 \msun and all separations
(Table \ref{tab:multiplicity_literature}).
This compilation is incomplete as we exclude surveys that only report
multiplicity for masses below 1 \msun and those without CSF information.
The above high-mass limit is set by the available multiplicity surveys.
and the low-mass cutoff is chosen as a reasonable detection threshold for stars in more
distant massive young clusters with both current (e.g. Keck) and
future (e.g. TMT) spectroscopy.
The compiled MF and CSF data (Figure \ref{fig:binary_properties}) are fit with a power-law
dependence on mass and give the following results:
\begin{eqnarray}
MF(m) = A m^{\gamma} \;\;\;\;\;\; A = 0.44 \;\;\; \gamma = 0.51 \label{eqn:multi_MF}
\\
CSF(m) = B m^{\beta} \;\;\;\;\;\; B = 0.50 \;\;\; \beta = 0.45 \label{eqn:multi_CSF}
\end{eqnarray}
The best-fit power laws are also shown in Figure \ref{fig:binary_properties}.
For any star cluster older than a few crossing times, the CSF will not continue to grow
indefinitely at high masses due to the dynamical instability of high-order systems in
a clustered environment.
A hard limit to the mean number of companions is adopted (CSF$\leq$3) such that systems
with more than 3 components are allowed, but have a low probability of occurrence.
Lastly, some observations suggest multiplicity properties may evolve over time differently
for clusters with different masses or stellar densities
\citep[e.~g.][]{kroupa1995,duchene1999,kohler2006,marks2012}.
However, the completeness and significance of these results is still uncertain
\citep{king2012} and we therefore assume all young clusters ($<$10 Myr) have
the same multiplicity properties, including the Young Nuclear Cluster.

\begin{deluxetable*}{rrrl}
\tablewidth{0pt}
\tablecaption{Multiplicity Measurements for Young Clusters}
\tablehead{
\colhead{Mass (\msun)} &
\colhead{MF} &
\colhead{CSF} &
\colhead{Reference}
}
\startdata
0.175 & 0.16 & 0.16 & Cha I, \citet{lafreniere2008} \\
0.390 & 0.33 & 0.37 & Cha I, \citet{lafreniere2008} \\
0.915 & 0.38 & 0.50 & Cha I, \citet{lafreniere2008} \\
2.14  & 0.63 & 0.75 & Cha I, \citet{lafreniere2008} \\
3.4   & 0.85 & -    & Sco OB2, \citet{kouwenhoven2007}\tablenotemark{a} \\
12.7  & 1.00 & 1.5  & Orion (ONC), \citet{preibisch1999} \\
16.8  & 1.00 & -    & Cyg OB2, \citet{kobulnickyFryer2007}\tablenotemark{b} \\
\enddata
\label{tab:multiplicity_literature}
\tablenotetext{a}{Mean mass of the sample was estimated from \citet{rizzuto2011}}
\tablenotetext{b}{Mean mass of the sample was estimated from \citet{kiminki2007}}

\end{deluxetable*}

{\it Facilities:} \facility{Keck: II (NIRC2)}, \facility{Keck: II (OSIRIS)}

\end{document}